\documentclass[10pt,english,aps,prl,twocolumn,superscriptaddress]{revtex4-2}
\usepackage{cancel}
\usepackage{mathtools}
\usepackage[T1]{fontenc}
\usepackage{graphicx}% Include figure files
\usepackage{bbm}% bold math
\usepackage[usenames,dvipsnames]{xcolor}
\usepackage{float}
\usepackage{stmaryrd}
\usepackage{tabularx}
\usepackage{nicematrix}
\usepackage{epstopdf}

\usepackage{amsmath} 
\usepackage[colorlinks=true,citecolor=blue,linkcolor=RubineRed,urlcolor=blue]{hyperref}

\begin{document}

\title{Non-Hermitian ramp in the interacting Hatano-Nelson Model}
\title{Slow approach to adiabaticity in many-body non-Hermitian systems:\\ the Hatano-Nelson Model}

\author{L\'eonce Dupays \href{https://orcid.org/0000-0002-3450-1861}{\includegraphics[scale=0.05]{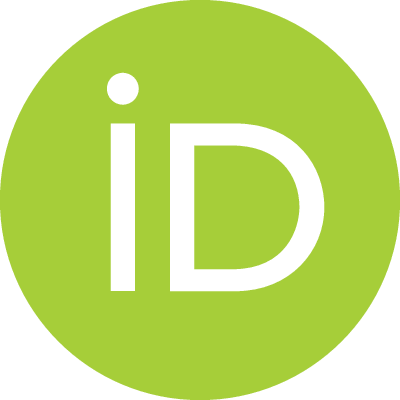}}}
\email{leonce.dupays@gmail.com}
\affiliation{Department  of  Physics  and  Materials  Science,  University  of  Luxembourg,  L-1511  Luxembourg,  Luxembourg}
%\affiliation{Department of Mathematics, King’s College London, Strand, London WC2R 2LS, UK}
\author{Adolfo del Campo\href{https://orcid.org/0000-0003-2219-2851}{\includegraphics[scale=0.05]{orcidid.eps}}}
\email{adolfo.delcampo@uni.lu}
\affiliation{Department  of  Physics  and  Materials  Science,  University  of  Luxembourg,  L-1511  Luxembourg,  Luxembourg}
\affiliation{Donostia International Physics Center,  E-20018 San Sebasti\'an, Spain}
\author{Bal\'azs D\'ora \href{https://orcid.org/0000-0002-5984-3761}{\includegraphics[scale=0.05]{orcidid.eps}}}
\email{dora.balazs@ttk.bme.hu}
\affiliation{Department of Theoretical Physics, Institute of Physics, Budapest University of Technology and Economics, M\H uegyetem rkp. 3., H-1111
Budapest, Hungary}

\begin{abstract}
We explore the near adiabatic dynamics in a non-Hermitian quantum many-body system by investigating a finite-time ramp of the imaginary vector potential in 
the interacting Hatano-Nelson model. The excess energy, the Loschmidt echo, and the density imbalance are analyzed using bosonization and exact diagonalization. 
The energy becomes complex valued, despite
the instantaneous Hamiltonian having the same real spectrum throughout. 
The adiabatic limit is approached very slowly through damped oscillations. The decay scales with $\tau^{-1}$ with $\tau$ the ramp duration,
 while the oscillation period is
$2L/v$ with $v$ the Fermi velocity and $L$ the system length. Yet, without the need for auxiliary controls, a shortcut to adiabaticity is found for ramp times commensurate with the period. 
Our work highlights the intricate interplay of adiabaticity and non-Hermitian many-body physics.

\end{abstract}
\maketitle

%\paragraph{Introduction.---} 
Adiabatic processes play a prominent role in various branches of physics \cite{beck}. In classical thermodynamics, adiabaticity is associated with the lack of heat transfer between the system and its environment and finds application in thermodynamic cycles as well as gas turbines, engines, and
compressors.
In a quantum setting \cite{messiah}, adiabaticity refers to the slow evolution of a quantum state such that the system remains in its instantaneous eigenstate \cite{polkovnikovnatphys,vignale}. This fundamental concept is relevant, e.g.,  for the Born–Oppenheimer approximation \cite{BO27,chruscinski2004geometric}, state preparation, and
adiabatic quantum computation \cite{nielsen}. 
Adiabatic protocols, well-studied in conventional Hermitian quantum mechanics, face new challenges and intriguing features when extended to non-Hermitian and open quantum
systems due to the presence of the environment.

Non-Hermitian systems \cite{ashidareview,Bergholtz2021, Brody_2014, Bender2007,MatsoukasRoubeas2023} arise in various physical contexts such as monitored quantum systems conditioned to measurement outcomes \cite{carmichael,daley,Jordan2024}, unitary evolution conditioned to a subspace via projection methods \cite{Muga04}, open quantum systems exchanging
energy or matter with an external environment, and in systems with gain and loss, which are commonly described by
complex potentials \cite{Muga04}. The Hamiltonians of such systems are non-Hermitian, generally leading to complex eigenvalues and non-orthogonal
eigenstates \cite{rotter,gao2015,zhou18,zeuner,gongprx,lee2016,takasu,fruchart,turkeshi,legal,lee2022,kunst2018,gongprx,kawabata,ElGanainy2018,russellyang}. Despite these features, non-Hermitian systems can exhibit real eigenvalues and phase transitions around exceptional
 points \cite{heiss,hodaei,ding2022}, where eigenvalues and eigenstates coalesce.
In non-Hermitian systems, the adiabatic theorem thus needs careful reformulation \cite{Garrison88,Ibanez14,Fleischer18}. 
%The key challenge lies in the non-orthogonality
%of eigenstates and the possibility of complex eigenvalues.

The prevalence of adiabaticity in non-Hermitian systems \cite{bender2007prl,Ibanez14} has profound implications for various fields, including photonics \cite{Feng2017}, where it can guide the
design of optical devices with tailored gain and loss profiles \cite{ding2022}. In quantum computing, understanding adiabatic processes in
non-Hermitian systems can lead to new approaches to quantum state preparation and manipulation \cite{Motta2019}. Moreover, exploring non-Hermitian
dynamics opens avenues for controlling quantum many-body systems in regimes where traditional Hermitian assumptions do not hold and where additional speedups may be possible \cite{delcampo13,Hornedal2024}. 
The quest for fast driving protocols for the preparation of a target state without the requirement of slow driving has led to the development of shortcuts to adiabaticity for unitary and nonunitary dynamics \cite{Vacanti2014,Dann19,Alipour2020,Dupays2020,Dupays2021shortcutstosqueezed,Yin2022}, including non-Hermitian systems \cite{Ibanez11,Torosov13,Alipour2020}. However, progress to date has been restricted to single-particle systems.

Here, we study near-adiabatic dynamics in a paradigmatic many-body non-Hermitian system, the Hatano-Nelson model \cite{hatanonelson1,hatanonelson2}, which describes
fermions in the presence of an imaginary vector potential.
By ramping up the imaginary vector linearly in time, the evolution of the system is monitored by focusing on the excess energy, Loschmidt-echo and particle imbalance.
We find using a variety of methods that the adiabatic limit is approached very slowly through damped oscillations compared to Hermitian systems \cite{polkovnikovnatphys}.
The decay scales with $\tau^{-1}$ with $\tau$ the ramp duration,
 while the oscillation period is
$2L/v$ with $v$ the Fermi velocity and $L$ the system length. For finite ramp times commensurate with the period, the system periodically reaches its adiabatic state. This constitutes a realization of a shortcut to adiabaticity in a many-body non-Hermitian setting.

\paragraph{Hatano-Nelson model with time-dependent imaginary vector potential.}

The lattice realization of the Hatano-Nelson model \cite{hatanonelson2,hatanonelson1} consists of fermions hopping in one dimension in the presence of an imaginary vector potential. Experimentally, the imaginary vector potential and the Hatano-Nelson model has been realized using digital quantum computer \cite{shen2023} or cold atomic systems \cite{takasu,ren2022,guo2020,liang2022}. 
Its
interacting many-body Hamiltonian reads
\begin{gather}
H_{HN}(t)=\sum_{n=1}^{N-1} \frac J2 \exp(ah(t))c^\dagger_nc_{n+1}+\nonumber\\
+\frac J2 \exp(-ah(t)) c^\dagger_{n+1}c_n+U c^\dagger_nc_{n}c^\dagger_{n+1}c_{n+1},
\label{hamiltontb}
\end{gather}
where $J>0$ is the uniform hopping, while $h(t)$, $a$ and $U$ denote the time-dependent
imaginary vector potential, the lattice constant, and the nearest-neighbor interaction between particles, respectively. The interaction satisfies $|U|\leq J$ in order to have gapless excitation spectrum.
$N$ is the total number of lattice sites. We consider
open boundary condition (OBC) and half filling ($N/2$ particles). 
The model is studied numerically using exact diagonalization (ED) by ramping up the imaginary vector potential term, which is modeled as a
sequence of infinitesimal sudden
quenches. The linear ramp is discretized into 100-1000 steps, and the convergence of the numerics is checked. Possible experimental realizations are expected to follow the same discretization protocol.

The effective low-energy Hamiltonian of Eq. \eqref{hamiltontb} \cite{giamarchi,cazalillaboson,nersesyan,Hofstetter2004} is given by
\begin{gather}
H(t)=\int_0^L \frac{dx}{2\pi} v\left[K(\pi\Pi(x)-ih(t))^2+\frac 1K (\partial_x\phi(x))^2\right],
\label{hamboson}
\end{gather}
where $\Pi(x)$ and $\phi(x)$ are the dual fields satisfying
the regular commutation relation  \cite{cazalillaboson}, the imaginary vector potential $h(t)$  is explicitly time-dependent, $v$ is the Fermi velocity and $K$ denotes the Luttinger liquid
parameter, which is $K=(\pi/2)/[\pi-\arccos(U/J)]$ for the tight binding model. Other non-hermitian incarnations of Luttinger liquid physics were discussed in Ref. \cite{yamamoto}.
For open boundary conditions (OBC), the mode expansion of the fields is given by
\begin{subequations}
\begin{gather}
 \phi(x)=i\sum_{q>0}\sqrt{\frac{\pi K}{qL}}\sin(qx)\left[b_q-b^\dagger_q\right]
\label{phix},\\
\Theta(x)=\sum_{q>0}\sqrt{\frac{\pi}{KqL}}\cos(qx)\left[b_q+b^\dagger_q\right],
\label{thetax}
\end{gather}
\end{subequations}
with $\Pi(x)=\partial_x\Theta(x)/\pi$ and  the mode quantization $q=l\pi/L$ with $l=1,2,\dots$.
Using these, the canonical bosonic form of the above Hamiltonian reduces to
\begin{gather}
H(t)=\sum_{q>0}\omega(q)b^\dagger_qb_q+\sum_{q>0}ig_q(t)(b_q+b^+_q)+E_0(t),
\label{hboson}
\end{gather}
where $g_q(t)=h(t)v\sqrt K (1-\cos(qL)/\sqrt{\pi L q}$, $\omega(q)=vq$ and $E_0(t)=-vKLh(t)^2/2\pi$.

We note that the instantaneous spectrum of both $H_{HN}(t)$ in Eq. \eqref{hamiltontb} and that of $H(t)$ in Eq. \eqref{hboson} remains unchanged for any fixed imaginary vector potential $h(t)$ and is identical to the  $h=0$ case for OBC. This follows from a similarity transformation connecting the  Hermitian Hamiltonian with $h=0$ and the non-Hermitian one with $h\neq 0$ \cite{Dora2023PRB}. Therefore, during the time evolution, only the wavefunction changes but not the spectrum. 

To determine the time evolution of physical quantities in the presence of $h(t)$, one can try to write down the non-Hermitian Heisenberg equation of motion for the creation and
annihilation operators. However, due to non-Hermiticity, these equations of motion do not close \cite{schomerus} despite the quadratic nature of Eq. \eqref{hboson}.
Instead, we follow a different approach by explicitly writing down the system's time-dependent wavefunction. The ansatz we take is
\begin{gather}
|\Psi(t)\rangle=\prod_{q>0}\exp\left(i\beta_q(t)+\alpha_q(t) b^+_q\right)|0\rangle,
%=\nonumber\\
%=\prod_{q>0}\exp\left(i\beta_q(t)\right)\sum_n \frac{(\alpha_q(t))^n}%{\sqrt{n!}}|n\rangle_q,
\label{psit}
\end{gather}
where $|0\rangle$ is the bosonic vacuum and we start from $h(0)=0$, which gives $\beta_q(0)=\alpha_q(0)=0$. 
%Here, $|n\rangle_q$ is the occupation number state for the momentum $q$ sector containing $n$ bosons.
By plugging the coherent state, $|\Psi(t)\rangle$ in Eq. \eqref{psit} into the time-dependent non-Hermitian Schr\"odinger equation, $i\partial_t|\Psi(t)\rangle=H(t)|\Psi(t)\rangle$, and using $\exp(\alpha_q(t) b^+_q)|0\rangle =\sum_{n=0}^\infty \frac{(\alpha_q(t))^n}{\sqrt{n!}}|n\rangle_q$ with  $|n\rangle_q$ being the occupation number state for the momentum $q$ sector containing $n$ bosons,
we get for a given $q$ mode in Eq. \eqref{psit}
\begin{align}
&\sum_n\left[-\partial_t \beta_q(t) (\alpha_q(t))^n+in(\alpha_q(t))^{n-1}\partial_t\alpha_q(t)\right]\frac{|n\rangle_q}{\sqrt{n!}}\nonumber\\
&=\sum_n ig_q(t)\frac{(\alpha_q(t))^n}{\sqrt{n!}}\left(\sqrt{n+1}|n+1\rangle_q+\sqrt{n}|n-1\rangle_q\right)\nonumber\\
&+\sum_n\left(\omega(q)n+E_0(t)\right)(\alpha_q(t))^n\frac{|n\rangle_q}{\sqrt{n!}}.
\end{align}
Upon rearranging terms in the sums, we are left with two coupled differential equations as
\begin{subequations}
\begin{gather}
i\partial_t\alpha_q(t)=\omega(q)\alpha_q(t)+ig_{q}(t),\\
\partial_t \beta_q(t)=-ig_{q}(t)\alpha_q(t)-E_0(t).
\end{gather}
\end{subequations}
Together with the initial condition $\alpha_q(0)=0$, its general solution is
\begin{gather}
\alpha_{q}(t)=\int_{0}^{t}\exp(i\omega(q)(s-t))g_q(s)ds,\label{solution_alpha}
\end{gather}
and similarly for $\beta_q(t)$, which is not needed for our purposes. This allows us to study any driving protocol of interest, e.g., a linear ramp or even time-periodic driving. In the following,
we focus on $h(t)=h_0t/\tau$ for $0\leq t \leq \tau$, leading to
\begin{gather}
\alpha_q(t)=g_q(\tau)\frac{1-i\omega(q)t-\exp(-i\omega(q)t)}{\omega(q)^2\tau}.\label{exact_alpha}
\end{gather}
Other ramp protocols are discussed in the Supplemental Material.

\paragraph{Excess Energy.} Under unitary evolution, the difference between the final mean energy and the adiabatic mean energy describes the excess energy in the process due to the nonadiabatic driving \cite{polkovnikovnatphys}. 
In the quenched Hatano-Nelson model, we thus consider the non-Hermitian generalization
\begin{align}
E(t)=\frac{\langle \Psi(t)|H(t)|\Psi(t)\rangle}{\langle \Psi(t)|\Psi(t)\rangle}-E_{\rm gs},
\end{align}
where $E_{\rm gs}$ is the ground state energy of the final Hamiltonian. %For the present model, this is also the ground state energy of the Hamiltonian at any time: 
The ground state energy does not change during the dynamics since the instantaneous Hamiltonians are related to each other by a similarity transformation \cite{hatanonelson1,Dora2023work}. Furthermore, the expectation value of the Hamiltonian in the normalized right eigenbasis basis used above leads to the same expectation value as in the normalized biorthogonal basis. The mean energy is complex-valued due to the non-Hermitian Hamiltonian, albeit the Hamiltonian $H(t)$ possesses a real instantaneous spectrum. 
For an arbitrary quench, the excess energy only depends on the parameter $\alpha_{q}(t)$ as
\begin{align}
E(t)&=\sum_{q>0}vq|\alpha_{q}(t)|^{2}+\sum_{q>0}2ig_q(t)\mathcal{R}e(\alpha_{q}(t))+E_0(t).
\label{heating1}
\end{align}
In particular, for a linear quench, by using  the exact expression in Eq. \eqref{exact_alpha}, the  excess energy is computed analytically in terms of the polylogarithm functions  \cite{gradstein} ${\rm Li_p} (z)$ as
\begin{figure}
\includegraphics[width=0.45\textwidth]{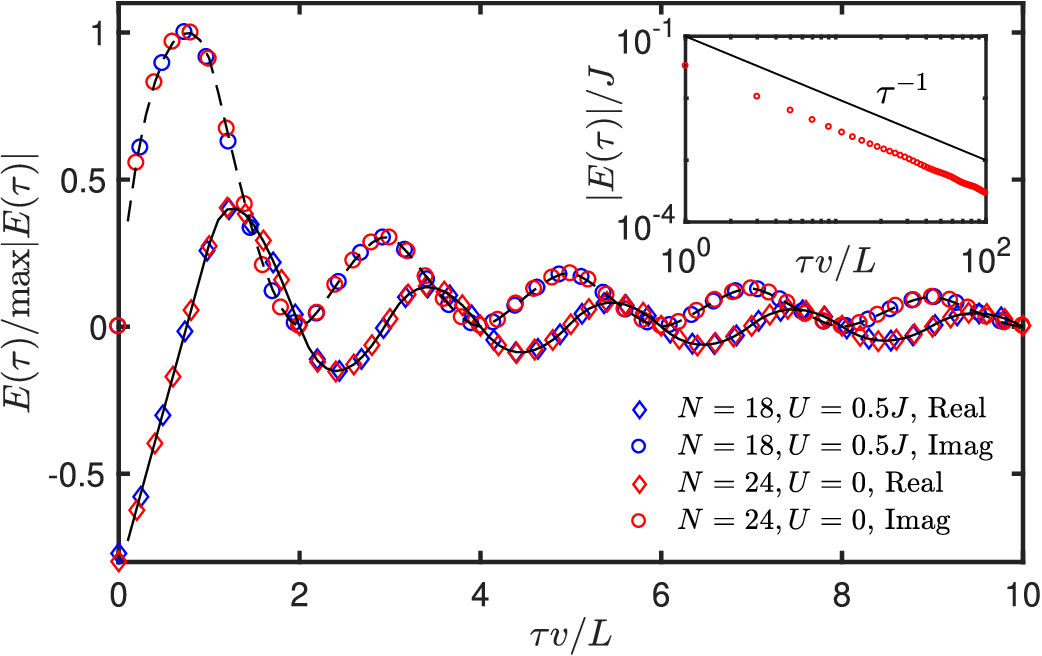}
\caption{Excess energy $E(\tau)$ for the quenched Hatano-Nelson model at the end of the ramp.
The numerical data from ED is displayed with symbols, while bosonization results are shown by  solid and dashed black lines for the real and imaginary parts, 
respectively.
In the adiabatic limit $\tau\rightarrow \infty$, the excess energy vanishes. There is a close agreement between the bosonization and numerics using $L=a(N+1)$ and $a h_{0}=0.1$. 
In addition, we have $N=24$, $U=0$ and $N=18$, $U=0.5J$. The inset highlights the $1/\tau$ decay (black solid lines) of the absolute value of the excess energy for $N=18$ and $U=0$, taken at the maxima from each oscillation. Note the extremely long final ramp times.  \label{fig:mean_energy}}
\end{figure}
\begin{align}
\frac{E({\tau})}{|E_0({\tau})|}&=\left(\frac{1}{\tilde{\tau}}\right)^{2}\bigg[-\frac{8}{\pi^{2}}\tilde{\tau}\sum_{\sigma=\pm}\mathcal{I}m\text{Li}_{3}(\sigma e^{i\sigma\tilde{\tau}})\nonumber\\
&-\frac{8}{\pi^{2}}\sum_{\sigma=\pm}\sigma \mathcal{R}e\text{Li}_{4}(\sigma e^{i\tilde{\tau}})+\frac{2\pi^{2}}{12}+\tilde{\tau}^{2}\bigg]\nonumber\\
&+\left(\frac{1}{\tilde{\tau}}\right)\frac{4i}{\pi^{2}}\bigg[\frac{7}{2}\zeta(3)-2\sum_{\sigma=\pm}\sigma\mathcal{R}e\text{Li}_{3}(\sigma e^{i\tilde{\tau}})\bigg]-1, \label{eq:mean_energy}
\end{align}
with the dimensionless time $\tilde{\tau}=(\tau  \pi v)/L$. Near the adiabatic limit $\tilde{\tau}\gg 1$, we find
\begin{align}
\frac{E({\tau})}{|E_0({\tau})|}&=\left(\frac{1}{\tilde{\tau}}\right)\Big\{-\frac{8}{\pi^{2}}\sum_{\sigma=\pm}\mathcal{I}m\text{Li}_{3}(\sigma e^{i\sigma\tilde{\tau}})\nonumber\\
&+\frac{4i}{\pi^{2}}\bigg[\frac{7}{2}\zeta(3)-2\sum_{\sigma=\pm}\sigma\mathcal{R}e\text{Li}_{3}(\sigma e^{i\tilde{\tau}})\bigg]\Big\},
\end{align}
showing a slow decay towards zero as $\tau^{-1}$ that is modulated by oscillations of period $2L/v$ for the real part and the imaginary part. 
This is in contrast to the case of Hermitian vector potential, where the excess energy decays faster as $\tau^{-2}$ \cite{polkovnikovnatphys}. 
In the former non-Hermitian case, the imaginary vector-potential dependent and independent terms in Eq. \eqref{hboson} give rise to imaginary and real energy expectation values in Eq. \eqref{heating1}, 
respectively, both decaying as $\tau^{-1}$. In the Hermitian case, on the other hand, both terms are real and still decay as $\tau^{-1}$ as a sort of equipartition theorem but cancel each other to leading order in $\tau$ as they can exchange energy. This results in a faster, $\tau^{-2}$  decay.

This slow approach to adiabaticity arises from a combination of the non-hermitian skin effect and the quantum Zeno effect. In contrast to Hermitian dynamics, the spatially homogeneous imaginary vector potential produces a density imbalance. Therefore, the redistribution of energy among particles slows down due to this inhomogenizing process. 
The quantum Zeno effect\cite{misra,ashidareview} from the imaginary vector potential slows down the propagation of correlations. In addition, the oscillations in $E(\tau)$ arise from OBC, representing a trapping box potential with characteristic frequency $v/L$, while their damping is due to the imaginary vector potential, causing the skin effect.

The maxima of the real part are reached for odd integers at times $\tau=(n+1/2)L/v$, and the minima for even integers. By contrast, the maxima of the imaginary part are reached for odd integer times $\tau=nL/v$ while minima are reached at even integer times. In the sudden quench limit, the energy is simply shifted 
${\rm lim}_{\tau\rightarrow 0}E(\tau)=-\frac{vh^{2}_{0}KL}{2\pi}=E_0(\tau)$.
Figure \ref{fig:mean_energy} depicts the excess energy after the quench at time $\tau$ and compares the bosonization result with the numerical simulation.  The excess energy is complex-valued, and there is a perfect agreement between the numerics and the analytical result. 

A prominent feature is the occurrence of a shortcut to adiabaticity for ramp times, which are integer multiples of $2L/v$. At such specific instances of time evolution, non-adiabatic excitations are exactly canceled out. At variance with common techniques for fast control, this is achieved without auxiliary fields. As a result, these driving protocols generalize to the non-Hermitian many-body setting the   ``accidental'' shortcut protocols known in the single-particle Hermitian case  \cite{Jaramillo_2016,Dupays2024exact,Dupays2024transitionless}.
\begin{figure}
    \centering
    \includegraphics[width=0.45\textwidth]{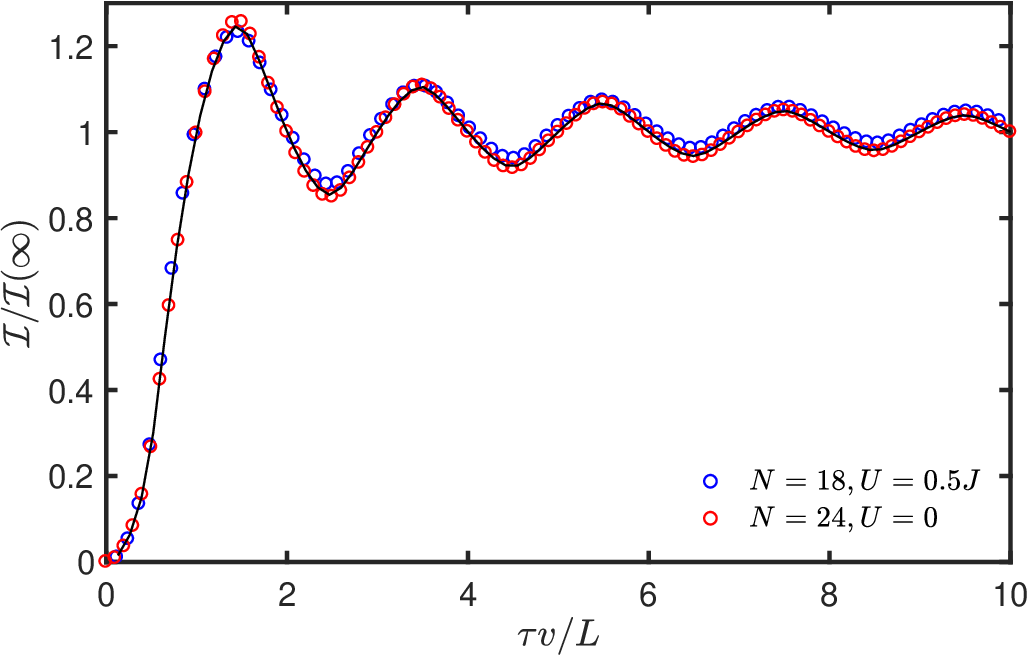}
    \caption{Evolution of the imbalance as a function of the quench duration from ED (symbols) and bosonization (line) for  $a h_{0}=0.1$. The imbalance reaches its asymptotic value at the integer times $2L/v$. }
    \label{fig:imbalance}
\end{figure}
Stopping the evolution at these times yields the adiabatic mean energy, as in truly adiabatic driving, but in finite time. 

%The excess energy during the quench is computed in the supplementary material (SM). The transitory regime exhibits a similar behavior to the final residual energy with an oscillatory behavior of the real and imaginary parts. Under the time rescaling $\tilde{t}=\pi t v/L$, the excess energy is independent of the quench duration with a $\tau^{-2}$ amplitude decay.
%
\paragraph{Imbalance.} 
As a proxy of the non-Hermitian skin effect, we focus on the imbalance  \cite{alsallom} between the two parts of the chain defined as
\begin{gather}
\mathcal{I}(t)=\frac{2}{N}\frac{\langle \Psi(t)|\left(\sum_{n\leq N/2}c^\dagger_{n}c_n-\sum_{N/2<n}c^\dagger_{n}c_n\right)|\Psi(t)\rangle}{\langle \Psi(t)|\Psi(t)\rangle }
\end{gather}
and the corresponding bosonized expression is
\begin{gather}
\mathcal{I}(t)=\frac{4}{L}\int_{0}^{L/2}\frac{\langle \Psi(t)|n_{0}(x)|\Psi(t)\rangle}{\langle \Psi(t)|\Psi(t)\rangle }dx=\nonumber\\
=\frac{4}{L\pi} \frac{\langle \Psi(t)|\phi(L/2)|\Psi(t)\rangle}{\langle \Psi(t)|\Psi(t)\rangle },
\end{gather}
which is related to the spatial average of the long wavelength density fluctuations $n_{0}(x,t)=\rho(x,t)-\rho_{0}=\partial_{x}\phi(x,t)/\pi$. This imbalance quantifies the asymmetry of the density distribution during the time evolution, which developes due to the skin effect. It can be directly computed using bosonization 
\begin{align}
\frac{\mathcal{I}(\tau)}{\mathcal{I}(\infty)}&=\frac{1}{\tilde{\tau}}\left\{\tilde{\tau}+\sum_{\sigma=\pm}\sigma\mathcal{R}e\text{Li}_{3}[\sigma i \exp(i\tilde{\tau})]\right\}, \label{eq:imbalance}
\end{align}
with the asymptotic value $\mathcal{I}(\infty)=16Kh_{0}\left(\frac{1}{\pi}\right)^{3} G$ with $G\approx 0.916$ the Catalan's constant and the initial value $\mathcal{I}(0)=0$.
We plot the imbalance in Fig. \ref{fig:imbalance} as a function of the quench time. 
At the initial time, the imbalance vanishes due to the homogeneous density of the LL. With increasing quench duration, the imbalance displays oscillations 
with period $2L/v$ around its asymptotic value $\mathcal{I}(\infty)$ with an amplitude decay in $\tau^{-1}$. At specific ramp times, which are integer multiples of $L/v$, the imbalance takes its adiabatic value. The imaginary vector potential drives the particles towards one of the boundaries due to skin effect, which then bounce back from the boundary, resulting in damped oscillations. 

This is to be contrasted with the case of a real vector potential, where no imbalance is expected in the adiabatic case. This follows from the fact that a real and homogeneous vector potential 
can be gauged away by a gauge transformation; thus, there is no particle imbalance in the adiabatic ground state in this case.
\begin{figure}
    \centering
    \includegraphics[width=0.45\textwidth]{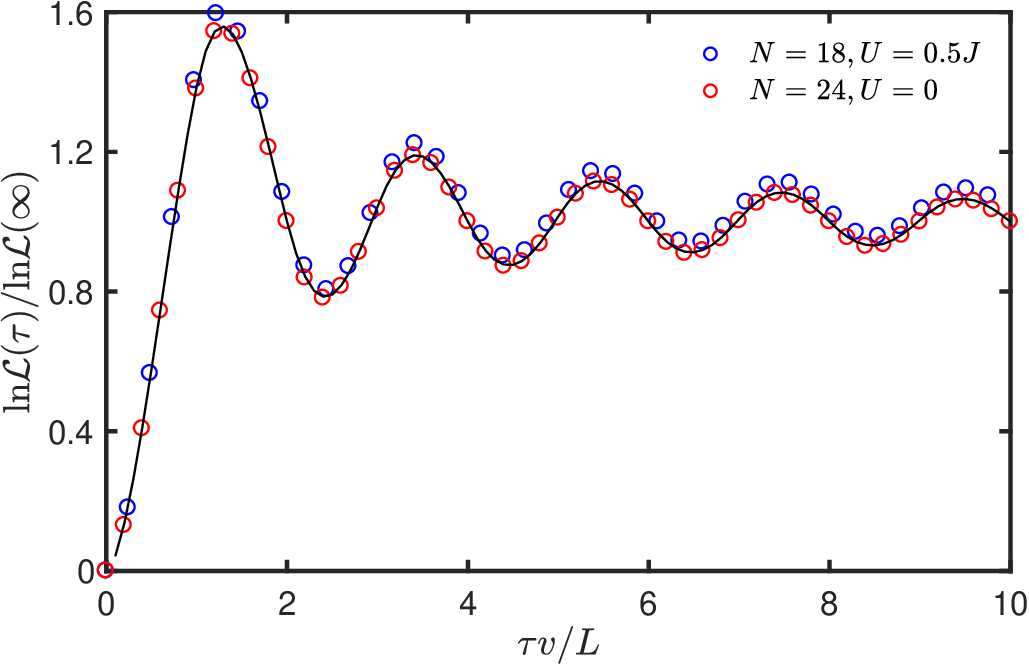}
    \caption{Evolution of the logarithm of the Loschmidt echo as a function of the quench duration. Comparison of the bosonized result, shown as a solid black line, with the numerical simulation (symbols), for $a h_{0}=0.1$.  }
    \label{fig:loschmidt}
\end{figure}
\paragraph{Loschmidt echo.} 
The Loschmidt echo \cite{silva,rmptalkner,goussev} or survival probability is the overlap of the final state with the initial one given by
\begin{gather}
\mathcal{L}(t)=\frac{\left|\langle 0|\Psi(t)\rangle\right|^2}{\langle\Psi(t)|\Psi(t)\rangle}=\exp\left(-\sum_{q>0}|\alpha_q(t)|^2\right).
\end{gather}
 It can be measured experimentally using Ramsey or single-qubit interferometry \cite{doraLE,Tonielli2020,Ness21}. We focus on the exponent of the Loschmidt echo ${\rm ln}\mathcal{L}(t)=-\sum_{q>0} |\alpha_{q}(t)|^{2}$ that is expressed analytically for the linear quench as
\begin{align}
&\frac{{\rm ln}\mathcal{L}(\tau)}{{\rm ln}\mathcal{L}(\infty)}=\left(\frac{1}{\tilde{\tau}}\right)^{2}\Big[\frac{31}{14}\frac{\zeta(5)}{\zeta(3)}-\frac{16}{14}\frac{1}{\zeta(3)}\sum_{\sigma=\pm}\sigma\mathcal{R}e\text{Li}_{5}\sigma e^{i\tilde{\tau}}\nonumber\\
&+\tilde{\tau}^{2}-\frac{8}{7\zeta(3)}\tilde{\tau}\sum_{\sigma=\pm}\sigma \mathcal{I}m \text{Li}_{4}\sigma e^{i\tilde{\tau}}\Big] \label{eq:loschmidt}
\end{align}
with the asymptotic value 
$\ln\mathcal{L}(\infty)=-\frac{7}{2}\frac{h^{2}_{0}KL^{2}}{\pi^{4}}\zeta(3)$, which is superextensive and grows with $L^2$, unlike the case of conventional many-body orthogonality catastrophe \cite{doraLE} in a LL, growing extensively as $\sim L$. At initial time ${\rm ln}\mathcal{L}(0)=0$. In the large time limit, the amplitude of the Loschmidt echo decays to leading order in $\tau^{-1}$ with 
oscillations of period $2L/v$ towards its asymptotic value. 

By inspecting the wavefunction at ramp times  integer multiples of  $2L/v$ from Eq. \eqref{exact_alpha}, we find 
$\alpha_q(\tau=2L/v)=-i g_q(\tau)/\omega(q)$, which is identical to not only the adiabatic expression but also agrees with the equilibrium wavefunction for
the imaginary vector potential $h_0$   \cite{Dora2023PRB}. Therefore, not only do the excess energy, the imbalance, and the Loschmidt echo take their adiabatic value for ramp times commensurate with $2L/v$, but all physical quantities match their adiabatic value identically due to this peculiar behavior of the wavefunction.
This is further verified numerically using ED by evaluating the overlap of the time-evolved wavefuncion with the equilibrium wavefunction for the imaginary vector potential $h_0$. The overlap is found to be unity within numerical accuracy for the above ramp times. 

\paragraph{Discussion and conclusion.} We have studied the approach to adiabaticity in a non-Hermitian gapless many-body system, by considering the paradigmatic Hatano-Nelson model quenched by an imaginary vector potential in finite time.  
For a linear quench, we find damped oscillatory behavior with a decaying amplitude in $\tau^{-1}$ for the excess energy, the imbalance that characterizes the asymmetry of the density distribution, and the Loschmidt echo. This decay is much slower than for the corresponding Hermitian vector potential quench, which is associated to the interplay of the non-hermitian skin effect and the quantum Zeno effect.
Our results show an excellent agreement between bosonization and numerical exact diagonalization already for relatively small systems.
We have further reported the exact cancellation of the 
residual mean energy at periodic quench times $\tau=2L/v$. This provides a shortcut to adiabaticity in a non-Hermitian many-body setting without the need to utilize auxiliary counterdiabatic controls.
Our findings open new avenues for control in quantum technologies and finite-time thermodynamics in non-Hermitian 
and open quantum systems.

\begin{acknowledgments}
This research was supported by the National Research, Development and Innovation Office - NKFIH  within the Quantum Technology National Excellence
Program (Project No.~2017-1.2.1-NKP-2017-00001), K134437, K142179 by the BME-Nanotechnology
FIKP grant (BME FIKP-NAT), and by a grant of the Ministry of Research, Innovation and
 Digitization, CNCS/CCCDI-UEFISCDI, under project number PN-III-P4-ID-PCE-2020-0277 and
 under the project for funding the excellence, Contract No. 29 PFE/30.12.2021. It was further supported by the Luxembourg National Research Fund (FNR), grant reference 17132054.
\end{acknowledgments}

\bibliography{Hatano,wboson1}

\begin{thebibliography}{76}%
\makeatletter
\providecommand \@ifxundefined [1]{%
 \@ifx{#1\undefined}
}%
\providecommand \@ifnum [1]{%
 \ifnum #1\expandafter \@firstoftwo
 \else \expandafter \@secondoftwo
 \fi
}%
\providecommand \@ifx [1]{%
 \ifx #1\expandafter \@firstoftwo
 \else \expandafter \@secondoftwo
 \fi
}%
\providecommand \natexlab [1]{#1}%
\providecommand \enquote  [1]{``#1''}%
\providecommand \bibnamefont  [1]{#1}%
\providecommand \bibfnamefont [1]{#1}%
\providecommand \citenamefont [1]{#1}%
\providecommand \href@noop [0]{\@secondoftwo}%
\providecommand \href [0]{\begingroup \@sanitize@url \@href}%
\providecommand \@href[1]{\@@startlink{#1}\@@href}%
\providecommand \@@href[1]{\endgroup#1\@@endlink}%
\providecommand \@sanitize@url [0]{\catcode `\\12\catcode `\$12\catcode
  `\&12\catcode `\#12\catcode `\^12\catcode `\_12\catcode `\%12\relax}%
\providecommand \@@startlink[1]{}%
\providecommand \@@endlink[0]{}%
\providecommand \url  [0]{\begingroup\@sanitize@url \@url }%
\providecommand \@url [1]{\endgroup\@href {#1}{\urlprefix }}%
\providecommand \urlprefix  [0]{URL }%
\providecommand \Eprint [0]{\href }%
\providecommand \doibase [0]{https://doi.org/}%
\providecommand \selectlanguage [0]{\@gobble}%
\providecommand \bibinfo  [0]{\@secondoftwo}%
\providecommand \bibfield  [0]{\@secondoftwo}%
\providecommand \translation [1]{[#1]}%
\providecommand \BibitemOpen [0]{}%
\providecommand \bibitemStop [0]{}%
\providecommand \bibitemNoStop [0]{.\EOS\space}%
\providecommand \EOS [0]{\spacefactor3000\relax}%
\providecommand \BibitemShut  [1]{\csname bibitem#1\endcsname}%
\let\auto@bib@innerbib\@empty
%</preamble>
\bibitem [{\citenamefont {Beck}\ and\ \citenamefont
  {Schl{\"o}gl}(1997)}]{beck}%
  \BibitemOpen
  \bibfield  {author} {\bibinfo {author} {\bibfnamefont {C.}~\bibnamefont
  {Beck}}\ and\ \bibinfo {author} {\bibfnamefont {F.}~\bibnamefont
  {Schl{\"o}gl}},\ }\href@noop {} {\emph {\bibinfo {title} {Thermodynamics of
  chaotic systems}}}\ (\bibinfo  {publisher} {Cambridge University Press},\
  \bibinfo {address} {New York},\ \bibinfo {year} {1997})\BibitemShut {NoStop}%
\bibitem [{\citenamefont {Messiah}(1961)}]{messiah}%
  \BibitemOpen
  \bibfield  {author} {\bibinfo {author} {\bibfnamefont {A.}~\bibnamefont
  {Messiah}},\ }\href@noop {} {\emph {\bibinfo {title} {Quantum Mechanics}}},\
  Dover books on physics\ (\bibinfo  {publisher} {Dover Publications},\
  \bibinfo {year} {1961})\BibitemShut {NoStop}%
\bibitem [{\citenamefont {Polkovnikov}\ and\ \citenamefont
  {Gritsev}(2008)}]{polkovnikovnatphys}%
  \BibitemOpen
  \bibfield  {author} {\bibinfo {author} {\bibfnamefont {A.}~\bibnamefont
  {Polkovnikov}}\ and\ \bibinfo {author} {\bibfnamefont {V.}~\bibnamefont
  {Gritsev}},\ }\bibfield  {title} {\bibinfo {title} {Breakdown of the
  adiabatic limit in low-dimensional gapless systems},\ }\href
  {https://doi.org/10.1038/nphys963} {\bibfield  {journal} {\bibinfo  {journal}
  {Nat. Phys}\ }\textbf {\bibinfo {volume} {4}},\ \bibinfo {pages} {477–481}
  (\bibinfo {year} {2008})}\BibitemShut {NoStop}%
\bibitem [{\citenamefont {Giuliani}\ and\ \citenamefont
  {Vignale}(2005)}]{vignale}%
  \BibitemOpen
  \bibfield  {author} {\bibinfo {author} {\bibfnamefont {G.}~\bibnamefont
  {Giuliani}}\ and\ \bibinfo {author} {\bibfnamefont {G.}~\bibnamefont
  {Vignale}},\ }\href@noop {} {\emph {\bibinfo {title} {Quantum Theory of the
  Electron Liquid}}}\ (\bibinfo  {publisher} {Cambridge University Press},\
  \bibinfo {address} {Cambridge},\ \bibinfo {year} {2005})\BibitemShut
  {NoStop}%
\bibitem [{\citenamefont {Born}\ and\ \citenamefont
  {Oppenheimer}(1927)}]{BO27}%
  \BibitemOpen
  \bibfield  {author} {\bibinfo {author} {\bibfnamefont {M.}~\bibnamefont
  {Born}}\ and\ \bibinfo {author} {\bibfnamefont {R.}~\bibnamefont
  {Oppenheimer}},\ }\bibfield  {title} {\bibinfo {title} {Zur quantentheorie
  der molekeln},\ }\href
  {https://doi.org/https://doi.org/10.1002/andp.19273892002} {\bibfield
  {journal} {\bibinfo  {journal} {Annalen der Physik}\ }\textbf {\bibinfo
  {volume} {389}},\ \bibinfo {pages} {457} (\bibinfo {year}
  {1927})}\BibitemShut {NoStop}%
\bibitem [{\citenamefont {Chruscinski}\ and\ \citenamefont
  {Jamiolkowski}(2004)}]{chruscinski2004geometric}%
  \BibitemOpen
  \bibfield  {author} {\bibinfo {author} {\bibfnamefont {D.}~\bibnamefont
  {Chruscinski}}\ and\ \bibinfo {author} {\bibfnamefont {A.}~\bibnamefont
  {Jamiolkowski}},\ }\href
  {https://doi.org/https://doi.org/10.1007/978-0-8176-8176-0} {\emph {\bibinfo
  {title} {Geometric Phases in Classical and Quantum Mechanics}}},\ Progress in
  Mathematical Physics\ (\bibinfo  {publisher} {Birkh{\"a}user Boston},\
  \bibinfo {year} {2004})\BibitemShut {NoStop}%
\bibitem [{\citenamefont {Nielsen}\ and\ \citenamefont
  {Chuang}(2000)}]{nielsen}%
  \BibitemOpen
  \bibfield  {author} {\bibinfo {author} {\bibfnamefont {M.}~\bibnamefont
  {Nielsen}}\ and\ \bibinfo {author} {\bibfnamefont {I.}~\bibnamefont
  {Chuang}},\ }\href@noop {} {\emph {\bibinfo {title} {Quantum Computation and
  Quantum Information}}}\ (\bibinfo  {publisher} {Cambridge University Press},\
  \bibinfo {address} {Cambridge},\ \bibinfo {year} {2000})\BibitemShut
  {NoStop}%
\bibitem [{\citenamefont {Ashida}\ \emph {et~al.}(2020)\citenamefont {Ashida},
  \citenamefont {Gong},\ and\ \citenamefont {Ueda}}]{ashidareview}%
  \BibitemOpen
  \bibfield  {author} {\bibinfo {author} {\bibfnamefont {Y.}~\bibnamefont
  {Ashida}}, \bibinfo {author} {\bibfnamefont {Z.}~\bibnamefont {Gong}},\ and\
  \bibinfo {author} {\bibfnamefont {M.}~\bibnamefont {Ueda}},\ }\bibfield
  {title} {\bibinfo {title} {Non-hermitian physics},\ }\href
  {https://doi.org/10.1080/00018732.2021.1876991} {\bibfield  {journal}
  {\bibinfo  {journal} {Adv. Phys.}\ }\textbf {\bibinfo {volume} {69}},\
  \bibinfo {pages} {249–435} (\bibinfo {year} {2020})}\BibitemShut {NoStop}%
\bibitem [{\citenamefont {Bergholtz}\ \emph {et~al.}(2021)\citenamefont
  {Bergholtz}, \citenamefont {Budich},\ and\ \citenamefont
  {Kunst}}]{Bergholtz2021}%
  \BibitemOpen
  \bibfield  {author} {\bibinfo {author} {\bibfnamefont {E.~J.}\ \bibnamefont
  {Bergholtz}}, \bibinfo {author} {\bibfnamefont {J.~C.}\ \bibnamefont
  {Budich}},\ and\ \bibinfo {author} {\bibfnamefont {F.~K.}\ \bibnamefont
  {Kunst}},\ }\bibfield  {title} {\bibinfo {title} {Exceptional topology of
  non-hermitian systems},\ }\href
  {https://doi.org/10.1103/RevModPhys.93.015005} {\bibfield  {journal}
  {\bibinfo  {journal} {Rev. Mod. Phys.}\ }\textbf {\bibinfo {volume} {93}},\
  \bibinfo {pages} {015005} (\bibinfo {year} {2021})}\BibitemShut {NoStop}%
\bibitem [{\citenamefont {Brody}(2013)}]{Brody_2014}%
  \BibitemOpen
  \bibfield  {author} {\bibinfo {author} {\bibfnamefont {D.~C.}\ \bibnamefont
  {Brody}},\ }\bibfield  {title} {\bibinfo {title} {Biorthogonal quantum
  mechanics},\ }\href {https://doi.org/10.1088/1751-8113/47/3/035305}
  {\bibfield  {journal} {\bibinfo  {journal} {J. Phys. A}\ }\textbf {\bibinfo
  {volume} {47}},\ \bibinfo {pages} {035305} (\bibinfo {year}
  {2013})}\BibitemShut {NoStop}%
\bibitem [{\citenamefont {Bender}(2007)}]{Bender2007}%
  \BibitemOpen
  \bibfield  {author} {\bibinfo {author} {\bibfnamefont {C.~M.}\ \bibnamefont
  {Bender}},\ }\bibfield  {title} {\bibinfo {title} {Making sense of
  non-hermitian hamiltonians},\ }\href
  {https://doi.org/10.1088/0034-4885/70/6/r03} {\bibfield  {journal} {\bibinfo
  {journal} {Rep. Prog. Phys.}\ }\textbf {\bibinfo {volume} {70}},\ \bibinfo
  {pages} {947} (\bibinfo {year} {2007})}\BibitemShut {NoStop}%
\bibitem [{\citenamefont {Matsoukas-Roubeas}\ \emph {et~al.}(2023)\citenamefont
  {Matsoukas-Roubeas}, \citenamefont {Roccati}, \citenamefont {Cornelius},
  \citenamefont {Xu}, \citenamefont {Chenu},\ and\ \citenamefont {del
  Campo}}]{MatsoukasRoubeas2023}%
  \BibitemOpen
  \bibfield  {author} {\bibinfo {author} {\bibfnamefont {A.~S.}\ \bibnamefont
  {Matsoukas-Roubeas}}, \bibinfo {author} {\bibfnamefont {F.}~\bibnamefont
  {Roccati}}, \bibinfo {author} {\bibfnamefont {J.}~\bibnamefont {Cornelius}},
  \bibinfo {author} {\bibfnamefont {Z.}~\bibnamefont {Xu}}, \bibinfo {author}
  {\bibfnamefont {A.}~\bibnamefont {Chenu}},\ and\ \bibinfo {author}
  {\bibfnamefont {A.}~\bibnamefont {del Campo}},\ }\bibfield  {title} {\bibinfo
  {title} {Non-hermitian hamiltonian deformations in quantum mechanics},\
  }\href {https://doi.org/10.1007/JHEP01(2023)060} {\bibfield  {journal}
  {\bibinfo  {journal} {Journal of High Energy Physics}\ }\textbf {\bibinfo
  {volume} {2023}},\ \bibinfo {pages} {60} (\bibinfo {year}
  {2023})}\BibitemShut {NoStop}%
\bibitem [{\citenamefont {Carmichael}(1993)}]{carmichael}%
  \BibitemOpen
  \bibfield  {author} {\bibinfo {author} {\bibfnamefont {H.}~\bibnamefont
  {Carmichael}},\ }\href@noop {} {\emph {\bibinfo {title} {An Open Systems
  Approach to Quantum Optics}}}\ (\bibinfo  {publisher} {Springer-Verlag},\
  \bibinfo {address} {Berlin},\ \bibinfo {year} {1993})\BibitemShut {NoStop}%
\bibitem [{\citenamefont {Daley}(2014)}]{daley}%
  \BibitemOpen
  \bibfield  {author} {\bibinfo {author} {\bibfnamefont {A.~J.}\ \bibnamefont
  {Daley}},\ }\bibfield  {title} {\bibinfo {title} {Quantum trajectories and
  open many-body quantum systems},\ }\href
  {https://doi.org/10.1080/00018732.2014.933502} {\bibfield  {journal}
  {\bibinfo  {journal} {Adv. Phys.}\ }\textbf {\bibinfo {volume} {63}},\
  \bibinfo {pages} {77} (\bibinfo {year} {2014})}\BibitemShut {NoStop}%
\bibitem [{\citenamefont {Jordan}\ and\ \citenamefont
  {Siddiqi}(2024)}]{Jordan2024}%
  \BibitemOpen
  \bibfield  {author} {\bibinfo {author} {\bibfnamefont {A.}~\bibnamefont
  {Jordan}}\ and\ \bibinfo {author} {\bibfnamefont {I.}~\bibnamefont
  {Siddiqi}},\ }\href {https://books.google.com/books?id=MRrxEAAAQBAJ} {\emph
  {\bibinfo {title} {Quantum Measurement: Theory and Practice}}}\ (\bibinfo
  {publisher} {Cambridge University Press},\ \bibinfo {year}
  {2024})\BibitemShut {NoStop}%
\bibitem [{\citenamefont {Muga}\ \emph {et~al.}(2004)\citenamefont {Muga},
  \citenamefont {Palao}, \citenamefont {Navarro},\ and\ \citenamefont
  {Egusquiza}}]{Muga04}%
  \BibitemOpen
  \bibfield  {author} {\bibinfo {author} {\bibfnamefont {J.}~\bibnamefont
  {Muga}}, \bibinfo {author} {\bibfnamefont {J.}~\bibnamefont {Palao}},
  \bibinfo {author} {\bibfnamefont {B.}~\bibnamefont {Navarro}},\ and\ \bibinfo
  {author} {\bibfnamefont {I.}~\bibnamefont {Egusquiza}},\ }\bibfield  {title}
  {\bibinfo {title} {Complex absorbing potentials},\ }\href
  {https://doi.org/https://doi.org/10.1016/j.physrep.2004.03.002} {\bibfield
  {journal} {\bibinfo  {journal} {Phys. Rep.}\ }\textbf {\bibinfo {volume}
  {395}},\ \bibinfo {pages} {357} (\bibinfo {year} {2004})}\BibitemShut
  {NoStop}%
\bibitem [{\citenamefont {Rotter}\ and\ \citenamefont {Bird}(2015)}]{rotter}%
  \BibitemOpen
  \bibfield  {author} {\bibinfo {author} {\bibfnamefont {I.}~\bibnamefont
  {Rotter}}\ and\ \bibinfo {author} {\bibfnamefont {J.~P.}\ \bibnamefont
  {Bird}},\ }\bibfield  {title} {\bibinfo {title} {A review of progress in the
  physics of open quantum systems: theory and experiment},\ }\href
  {https://iopscience.iop.org/article/10.1088/0034-4885/78/11/114001}
  {\bibfield  {journal} {\bibinfo  {journal} {Rep. Prog. Phys.}\ }\textbf
  {\bibinfo {volume} {78}},\ \bibinfo {pages} {114001} (\bibinfo {year}
  {2015})}\BibitemShut {NoStop}%
\bibitem [{\citenamefont {Gao}\ \emph {et~al.}(2015)\citenamefont {Gao},
  \citenamefont {Estrecho}, \citenamefont {Bliokh}, \citenamefont {Liew},
  \citenamefont {Fraser}, \citenamefont {Brodbeck}, \citenamefont {Kamp},
  \citenamefont {Schneider}, \citenamefont {H{\"o}fling}, \citenamefont
  {Yamamoto}, \citenamefont {Nori}, \citenamefont {Kivshar}, \citenamefont
  {Truscott}, \citenamefont {Dall},\ and\ \citenamefont
  {Ostrovskaya}}]{gao2015}%
  \BibitemOpen
  \bibfield  {author} {\bibinfo {author} {\bibfnamefont {T.}~\bibnamefont
  {Gao}}, \bibinfo {author} {\bibfnamefont {E.}~\bibnamefont {Estrecho}},
  \bibinfo {author} {\bibfnamefont {K.~Y.}\ \bibnamefont {Bliokh}}, \bibinfo
  {author} {\bibfnamefont {T.~C.~H.}\ \bibnamefont {Liew}}, \bibinfo {author}
  {\bibfnamefont {M.~D.}\ \bibnamefont {Fraser}}, \bibinfo {author}
  {\bibfnamefont {S.}~\bibnamefont {Brodbeck}}, \bibinfo {author}
  {\bibfnamefont {M.}~\bibnamefont {Kamp}}, \bibinfo {author} {\bibfnamefont
  {C.}~\bibnamefont {Schneider}}, \bibinfo {author} {\bibfnamefont
  {S.}~\bibnamefont {H{\"o}fling}}, \bibinfo {author} {\bibfnamefont
  {Y.}~\bibnamefont {Yamamoto}}, \bibinfo {author} {\bibfnamefont
  {F.}~\bibnamefont {Nori}}, \bibinfo {author} {\bibfnamefont {Y.~S.}\
  \bibnamefont {Kivshar}}, \bibinfo {author} {\bibfnamefont {A.~G.}\
  \bibnamefont {Truscott}}, \bibinfo {author} {\bibfnamefont {R.~G.}\
  \bibnamefont {Dall}},\ and\ \bibinfo {author} {\bibfnamefont {E.~A.}\
  \bibnamefont {Ostrovskaya}},\ }\bibfield  {title} {\bibinfo {title}
  {Observation of non-hermitian degeneracies in a chaotic exciton-polariton
  billiard},\ }\href {https://doi.org/10.1038/nature15522} {\bibfield
  {journal} {\bibinfo  {journal} {Nature}\ }\textbf {\bibinfo {volume} {526}},\
  \bibinfo {pages} {554} (\bibinfo {year} {2015})}\BibitemShut {NoStop}%
\bibitem [{\citenamefont {Zhou}\ \emph {et~al.}(2018)\citenamefont {Zhou},
  \citenamefont {Wang}, \citenamefont {Wang},\ and\ \citenamefont
  {Gong}}]{zhou18}%
  \BibitemOpen
  \bibfield  {author} {\bibinfo {author} {\bibfnamefont {L.}~\bibnamefont
  {Zhou}}, \bibinfo {author} {\bibfnamefont {Q.-h.}\ \bibnamefont {Wang}},
  \bibinfo {author} {\bibfnamefont {H.}~\bibnamefont {Wang}},\ and\ \bibinfo
  {author} {\bibfnamefont {J.}~\bibnamefont {Gong}},\ }\bibfield  {title}
  {\bibinfo {title} {Dynamical quantum phase transitions in non-hermitian
  lattices},\ }\href {https://doi.org/10.1103/PhysRevA.98.022129} {\bibfield
  {journal} {\bibinfo  {journal} {Phys. Rev. A}\ }\textbf {\bibinfo {volume}
  {98}},\ \bibinfo {pages} {022129} (\bibinfo {year} {2018})}\BibitemShut
  {NoStop}%
\bibitem [{\citenamefont {Zeuner}\ \emph {et~al.}(2015)\citenamefont {Zeuner},
  \citenamefont {Rechtsman}, \citenamefont {Plotnik}, \citenamefont {Lumer},
  \citenamefont {Nolte}, \citenamefont {Rudner}, \citenamefont {Segev},\ and\
  \citenamefont {Szameit}}]{zeuner}%
  \BibitemOpen
  \bibfield  {author} {\bibinfo {author} {\bibfnamefont {J.~M.}\ \bibnamefont
  {Zeuner}}, \bibinfo {author} {\bibfnamefont {M.~C.}\ \bibnamefont
  {Rechtsman}}, \bibinfo {author} {\bibfnamefont {Y.}~\bibnamefont {Plotnik}},
  \bibinfo {author} {\bibfnamefont {Y.}~\bibnamefont {Lumer}}, \bibinfo
  {author} {\bibfnamefont {S.}~\bibnamefont {Nolte}}, \bibinfo {author}
  {\bibfnamefont {M.~S.}\ \bibnamefont {Rudner}}, \bibinfo {author}
  {\bibfnamefont {M.}~\bibnamefont {Segev}},\ and\ \bibinfo {author}
  {\bibfnamefont {A.}~\bibnamefont {Szameit}},\ }\bibfield  {title} {\bibinfo
  {title} {Observation of a topological transition in the bulk of a
  non-hermitian system},\ }\href
  {https://doi.org/10.1103/PhysRevLett.115.040402} {\bibfield  {journal}
  {\bibinfo  {journal} {Phys. Rev. Lett.}\ }\textbf {\bibinfo {volume} {115}},\
  \bibinfo {pages} {040402} (\bibinfo {year} {2015})}\BibitemShut {NoStop}%
\bibitem [{\citenamefont {Gong}\ \emph {et~al.}(2018)\citenamefont {Gong},
  \citenamefont {Ashida}, \citenamefont {Kawabata}, \citenamefont {Takasan},
  \citenamefont {Higashikawa},\ and\ \citenamefont {Ueda}}]{gongprx}%
  \BibitemOpen
  \bibfield  {author} {\bibinfo {author} {\bibfnamefont {Z.}~\bibnamefont
  {Gong}}, \bibinfo {author} {\bibfnamefont {Y.}~\bibnamefont {Ashida}},
  \bibinfo {author} {\bibfnamefont {K.}~\bibnamefont {Kawabata}}, \bibinfo
  {author} {\bibfnamefont {K.}~\bibnamefont {Takasan}}, \bibinfo {author}
  {\bibfnamefont {S.}~\bibnamefont {Higashikawa}},\ and\ \bibinfo {author}
  {\bibfnamefont {M.}~\bibnamefont {Ueda}},\ }\bibfield  {title} {\bibinfo
  {title} {Topological phases of non-hermitian systems},\ }\href
  {https://doi.org/10.1103/PhysRevX.8.031079} {\bibfield  {journal} {\bibinfo
  {journal} {Phys. Rev. X}\ }\textbf {\bibinfo {volume} {8}},\ \bibinfo {pages}
  {031079} (\bibinfo {year} {2018})}\BibitemShut {NoStop}%
\bibitem [{\citenamefont {Lee}(2016)}]{lee2016}%
  \BibitemOpen
  \bibfield  {author} {\bibinfo {author} {\bibfnamefont {T.~E.}\ \bibnamefont
  {Lee}},\ }\bibfield  {title} {\bibinfo {title} {Anomalous edge state in a
  non-hermitian lattice},\ }\href
  {https://doi.org/10.1103/PhysRevLett.116.133903} {\bibfield  {journal}
  {\bibinfo  {journal} {Phys. Rev. Lett.}\ }\textbf {\bibinfo {volume} {116}},\
  \bibinfo {pages} {133903} (\bibinfo {year} {2016})}\BibitemShut {NoStop}%
\bibitem [{\citenamefont {Takasu}\ \emph {et~al.}(2020)\citenamefont {Takasu},
  \citenamefont {Yagami}, \citenamefont {Ashida}, \citenamefont {Hamazaki},
  \citenamefont {Kuno},\ and\ \citenamefont {Takahashi}}]{takasu}%
  \BibitemOpen
  \bibfield  {author} {\bibinfo {author} {\bibfnamefont {Y.}~\bibnamefont
  {Takasu}}, \bibinfo {author} {\bibfnamefont {T.}~\bibnamefont {Yagami}},
  \bibinfo {author} {\bibfnamefont {Y.}~\bibnamefont {Ashida}}, \bibinfo
  {author} {\bibfnamefont {R.}~\bibnamefont {Hamazaki}}, \bibinfo {author}
  {\bibfnamefont {Y.}~\bibnamefont {Kuno}},\ and\ \bibinfo {author}
  {\bibfnamefont {Y.}~\bibnamefont {Takahashi}},\ }\bibfield  {title} {\bibinfo
  {title} {{PT-symmetric non-Hermitian quantum many-body system using ultracold
  atoms in an optical lattice with controlled dissipation}},\ }\href
  {https://doi.org/10.1093/ptep/ptaa094} {\bibfield  {journal} {\bibinfo
  {journal} {PTEP}\ } (\bibinfo {year} {2020})}\BibitemShut {NoStop}%
\bibitem [{\citenamefont {Fruchart}\ \emph {et~al.}(2021)\citenamefont
  {Fruchart}, \citenamefont {Hanai}, \citenamefont {Littlewood},\ and\
  \citenamefont {Vitelli}}]{fruchart}%
  \BibitemOpen
  \bibfield  {author} {\bibinfo {author} {\bibfnamefont {M.}~\bibnamefont
  {Fruchart}}, \bibinfo {author} {\bibfnamefont {R.}~\bibnamefont {Hanai}},
  \bibinfo {author} {\bibfnamefont {P.~B.}\ \bibnamefont {Littlewood}},\ and\
  \bibinfo {author} {\bibfnamefont {V.}~\bibnamefont {Vitelli}},\ }\bibfield
  {title} {\bibinfo {title} {Non-reciprocal phase transitions},\ }\href
  {https://doi.org/10.1038/s41586-021-03375-9} {\bibfield  {journal} {\bibinfo
  {journal} {Nature}\ }\textbf {\bibinfo {volume} {592}},\ \bibinfo {pages}
  {363} (\bibinfo {year} {2021})}\BibitemShut {NoStop}%
\bibitem [{\citenamefont {Turkeshi}\ and\ \citenamefont
  {Schir\'o}(2023)}]{turkeshi}%
  \BibitemOpen
  \bibfield  {author} {\bibinfo {author} {\bibfnamefont {X.}~\bibnamefont
  {Turkeshi}}\ and\ \bibinfo {author} {\bibfnamefont {M.}~\bibnamefont
  {Schir\'o}},\ }\bibfield  {title} {\bibinfo {title} {Entanglement and
  correlation spreading in non-hermitian spin chains},\ }\href
  {https://doi.org/10.1103/PhysRevB.107.L020403} {\bibfield  {journal}
  {\bibinfo  {journal} {Phys. Rev. B}\ }\textbf {\bibinfo {volume} {107}},\
  \bibinfo {pages} {L020403} (\bibinfo {year} {2023})}\BibitemShut {NoStop}%
\bibitem [{\citenamefont {Gal}\ \emph {et~al.}(2023)\citenamefont {Gal},
  \citenamefont {Turkeshi},\ and\ \citenamefont {Schirò}}]{legal}%
  \BibitemOpen
  \bibfield  {author} {\bibinfo {author} {\bibfnamefont {Y.~L.}\ \bibnamefont
  {Gal}}, \bibinfo {author} {\bibfnamefont {X.}~\bibnamefont {Turkeshi}},\ and\
  \bibinfo {author} {\bibfnamefont {M.}~\bibnamefont {Schirò}},\ }\bibfield
  {title} {\bibinfo {title} {{Volume-to-area law entanglement transition in a
  non-Hermitian free fermionic chain}},\ }\href
  {https://doi.org/10.21468/SciPostPhys.14.5.138} {\bibfield  {journal}
  {\bibinfo  {journal} {SciPost Phys.}\ }\textbf {\bibinfo {volume} {14}},\
  \bibinfo {pages} {138} (\bibinfo {year} {2023})}\BibitemShut {NoStop}%
\bibitem [{\citenamefont {Lee}(2022)}]{lee2022}%
  \BibitemOpen
  \bibfield  {author} {\bibinfo {author} {\bibfnamefont {C.~H.}\ \bibnamefont
  {Lee}},\ }\bibfield  {title} {\bibinfo {title} {Exceptional bound states and
  negative entanglement entropy},\ }\href
  {https://doi.org/10.1103/PhysRevLett.128.010402} {\bibfield  {journal}
  {\bibinfo  {journal} {Phys. Rev. Lett.}\ }\textbf {\bibinfo {volume} {128}},\
  \bibinfo {pages} {010402} (\bibinfo {year} {2022})}\BibitemShut {NoStop}%
\bibitem [{\citenamefont {Kunst}\ \emph {et~al.}(2018)\citenamefont {Kunst},
  \citenamefont {Edvardsson}, \citenamefont {Budich},\ and\ \citenamefont
  {Bergholtz}}]{kunst2018}%
  \BibitemOpen
  \bibfield  {author} {\bibinfo {author} {\bibfnamefont {F.~K.}\ \bibnamefont
  {Kunst}}, \bibinfo {author} {\bibfnamefont {E.}~\bibnamefont {Edvardsson}},
  \bibinfo {author} {\bibfnamefont {J.~C.}\ \bibnamefont {Budich}},\ and\
  \bibinfo {author} {\bibfnamefont {E.~J.}\ \bibnamefont {Bergholtz}},\
  }\bibfield  {title} {\bibinfo {title} {Biorthogonal bulk-boundary
  correspondence in non-hermitian systems},\ }\href
  {https://doi.org/10.1103/PhysRevLett.121.026808} {\bibfield  {journal}
  {\bibinfo  {journal} {Phys. Rev. Lett.}\ }\textbf {\bibinfo {volume} {121}},\
  \bibinfo {pages} {026808} (\bibinfo {year} {2018})}\BibitemShut {NoStop}%
\bibitem [{\citenamefont {Kawabata}\ \emph {et~al.}(2023)\citenamefont
  {Kawabata}, \citenamefont {Numasawa},\ and\ \citenamefont {Ryu}}]{kawabata}%
  \BibitemOpen
  \bibfield  {author} {\bibinfo {author} {\bibfnamefont {K.}~\bibnamefont
  {Kawabata}}, \bibinfo {author} {\bibfnamefont {T.}~\bibnamefont {Numasawa}},\
  and\ \bibinfo {author} {\bibfnamefont {S.}~\bibnamefont {Ryu}},\ }\bibfield
  {title} {\bibinfo {title} {Entanglement phase transition induced by the
  non-hermitian skin effect},\ }\href
  {https://doi.org/10.1103/PhysRevX.13.021007} {\bibfield  {journal} {\bibinfo
  {journal} {Phys. Rev. X}\ }\textbf {\bibinfo {volume} {13}},\ \bibinfo
  {pages} {021007} (\bibinfo {year} {2023})}\BibitemShut {NoStop}%
\bibitem [{\citenamefont {El-Ganainy}\ \emph {et~al.}(2018)\citenamefont
  {El-Ganainy}, \citenamefont {Makris}, \citenamefont {Khajavikhan},
  \citenamefont {Musslimani}, \citenamefont {Rotter},\ and\ \citenamefont
  {Christodoulides}}]{ElGanainy2018}%
  \BibitemOpen
  \bibfield  {author} {\bibinfo {author} {\bibfnamefont {R.}~\bibnamefont
  {El-Ganainy}}, \bibinfo {author} {\bibfnamefont {K.~G.}\ \bibnamefont
  {Makris}}, \bibinfo {author} {\bibfnamefont {M.}~\bibnamefont {Khajavikhan}},
  \bibinfo {author} {\bibfnamefont {Z.~H.}\ \bibnamefont {Musslimani}},
  \bibinfo {author} {\bibfnamefont {S.}~\bibnamefont {Rotter}},\ and\ \bibinfo
  {author} {\bibfnamefont {D.~N.}\ \bibnamefont {Christodoulides}},\ }\bibfield
   {title} {\bibinfo {title} {Non-hermitian physics and pt symmetry},\ }\href
  {https://doi.org/10.1038/nphys4323} {\bibfield  {journal} {\bibinfo
  {journal} {Nat. Phys.}\ }\textbf {\bibinfo {volume} {14}},\ \bibinfo {pages}
  {11} (\bibinfo {year} {2018})}\BibitemShut {NoStop}%
\bibitem [{\citenamefont {Yang}\ \emph {et~al.}(2022)\citenamefont {Yang},
  \citenamefont {Tan}, \citenamefont {Tai}, \citenamefont {Koh}, \citenamefont
  {Li}, \citenamefont {Longhi},\ and\ \citenamefont {Lee}}]{russellyang}%
  \BibitemOpen
  \bibfield  {author} {\bibinfo {author} {\bibfnamefont {R.}~\bibnamefont
  {Yang}}, \bibinfo {author} {\bibfnamefont {J.~W.}\ \bibnamefont {Tan}},
  \bibinfo {author} {\bibfnamefont {T.}~\bibnamefont {Tai}}, \bibinfo {author}
  {\bibfnamefont {J.~M.}\ \bibnamefont {Koh}}, \bibinfo {author} {\bibfnamefont
  {L.}~\bibnamefont {Li}}, \bibinfo {author} {\bibfnamefont {S.}~\bibnamefont
  {Longhi}},\ and\ \bibinfo {author} {\bibfnamefont {C.~H.}\ \bibnamefont
  {Lee}},\ }\bibfield  {title} {\bibinfo {title} {Designing non-hermitian real
  spectra through electrostatics},\ }\href
  {https://doi.org/https://doi.org/10.1016/j.scib.2022.08.005} {\bibfield
  {journal} {\bibinfo  {journal} {Science Bulletin}\ }\textbf {\bibinfo
  {volume} {67}},\ \bibinfo {pages} {1865} (\bibinfo {year}
  {2022})}\BibitemShut {NoStop}%
\bibitem [{\citenamefont {Heiss}(2012)}]{heiss}%
  \BibitemOpen
  \bibfield  {author} {\bibinfo {author} {\bibfnamefont {W.~D.}\ \bibnamefont
  {Heiss}},\ }\bibfield  {title} {\bibinfo {title} {The physics of exceptional
  points},\ }\href {https://doi.org/10.1088/1751-8113/45/44/444016} {\bibfield
  {journal} {\bibinfo  {journal} {J. Phys. A: Math. Theor.}\ }\textbf {\bibinfo
  {volume} {45}},\ \bibinfo {pages} {444016} (\bibinfo {year}
  {2012})}\BibitemShut {NoStop}%
\bibitem [{\citenamefont {Hodaei}\ \emph {et~al.}(2017)\citenamefont {Hodaei},
  \citenamefont {Hassan}, \citenamefont {Wittek}, \citenamefont
  {Garcia-Gracia}, \citenamefont {El-Ganainy}, \citenamefont
  {Christodoulides},\ and\ \citenamefont {Khajavikhan}}]{hodaei}%
  \BibitemOpen
  \bibfield  {author} {\bibinfo {author} {\bibfnamefont {H.}~\bibnamefont
  {Hodaei}}, \bibinfo {author} {\bibfnamefont {A.~U.}\ \bibnamefont {Hassan}},
  \bibinfo {author} {\bibfnamefont {S.}~\bibnamefont {Wittek}}, \bibinfo
  {author} {\bibfnamefont {H.}~\bibnamefont {Garcia-Gracia}}, \bibinfo {author}
  {\bibfnamefont {R.}~\bibnamefont {El-Ganainy}}, \bibinfo {author}
  {\bibfnamefont {D.~N.}\ \bibnamefont {Christodoulides}},\ and\ \bibinfo
  {author} {\bibfnamefont {M.}~\bibnamefont {Khajavikhan}},\ }\bibfield
  {title} {\bibinfo {title} {Enhanced sensitivity at higher-order exceptional
  points},\ }\href {https://doi.org/10.1038/nature23280} {\bibfield  {journal}
  {\bibinfo  {journal} {Nature}\ }\textbf {\bibinfo {volume} {548}},\ \bibinfo
  {pages} {187} (\bibinfo {year} {2017})}\BibitemShut {NoStop}%
\bibitem [{\citenamefont {Ding}\ \emph {et~al.}(2022)\citenamefont {Ding},
  \citenamefont {Fang},\ and\ \citenamefont {Ma}}]{ding2022}%
  \BibitemOpen
  \bibfield  {author} {\bibinfo {author} {\bibfnamefont {K.}~\bibnamefont
  {Ding}}, \bibinfo {author} {\bibfnamefont {C.}~\bibnamefont {Fang}},\ and\
  \bibinfo {author} {\bibfnamefont {G.}~\bibnamefont {Ma}},\ }\bibfield
  {title} {\bibinfo {title} {Non-hermitian topology and exceptional-point
  geometries},\ }\href {https://doi.org/10.1038/s42254-022-00516-5} {\bibfield
  {journal} {\bibinfo  {journal} {Nat. Rev. Phys.}\ }\textbf {\bibinfo {volume}
  {4}},\ \bibinfo {pages} {745} (\bibinfo {year} {2022})}\BibitemShut {NoStop}%
\bibitem [{\citenamefont {Garrison}\ and\ \citenamefont
  {Wright}(1988)}]{Garrison88}%
  \BibitemOpen
  \bibfield  {author} {\bibinfo {author} {\bibfnamefont {J.}~\bibnamefont
  {Garrison}}\ and\ \bibinfo {author} {\bibfnamefont {E.}~\bibnamefont
  {Wright}},\ }\bibfield  {title} {\bibinfo {title} {Complex geometrical phases
  for dissipative systems},\ }\href
  {https://doi.org/https://doi.org/10.1016/0375-9601(88)90905-X} {\bibfield
  {journal} {\bibinfo  {journal} {Physics Letters A}\ }\textbf {\bibinfo
  {volume} {128}},\ \bibinfo {pages} {177} (\bibinfo {year}
  {1988})}\BibitemShut {NoStop}%
\bibitem [{\citenamefont {Ib\'a\~nez}\ and\ \citenamefont
  {Muga}(2014)}]{Ibanez14}%
  \BibitemOpen
  \bibfield  {author} {\bibinfo {author} {\bibfnamefont {S.}~\bibnamefont
  {Ib\'a\~nez}}\ and\ \bibinfo {author} {\bibfnamefont {J.~G.}\ \bibnamefont
  {Muga}},\ }\bibfield  {title} {\bibinfo {title} {Adiabaticity condition for
  non-hermitian hamiltonians},\ }\href
  {https://doi.org/10.1103/PhysRevA.89.033403} {\bibfield  {journal} {\bibinfo
  {journal} {Phys. Rev. A}\ }\textbf {\bibinfo {volume} {89}},\ \bibinfo
  {pages} {033403} (\bibinfo {year} {2014})}\BibitemShut {NoStop}%
\bibitem [{\citenamefont {Fleischer}\ and\ \citenamefont
  {Moiseyev}(2005)}]{Fleischer18}%
  \BibitemOpen
  \bibfield  {author} {\bibinfo {author} {\bibfnamefont {A.}~\bibnamefont
  {Fleischer}}\ and\ \bibinfo {author} {\bibfnamefont {N.}~\bibnamefont
  {Moiseyev}},\ }\bibfield  {title} {\bibinfo {title} {Adiabatic theorem for
  non-hermitian time-dependent open systems},\ }\href
  {https://doi.org/10.1103/PhysRevA.72.032103} {\bibfield  {journal} {\bibinfo
  {journal} {Phys. Rev. A}\ }\textbf {\bibinfo {volume} {72}},\ \bibinfo
  {pages} {032103} (\bibinfo {year} {2005})}\BibitemShut {NoStop}%
\bibitem [{\citenamefont {Bender}\ \emph {et~al.}(2007)\citenamefont {Bender},
  \citenamefont {Brody}, \citenamefont {Jones},\ and\ \citenamefont
  {Meister}}]{bender2007prl}%
  \BibitemOpen
  \bibfield  {author} {\bibinfo {author} {\bibfnamefont {C.~M.}\ \bibnamefont
  {Bender}}, \bibinfo {author} {\bibfnamefont {D.~C.}\ \bibnamefont {Brody}},
  \bibinfo {author} {\bibfnamefont {H.~F.}\ \bibnamefont {Jones}},\ and\
  \bibinfo {author} {\bibfnamefont {B.~K.}\ \bibnamefont {Meister}},\
  }\bibfield  {title} {\bibinfo {title} {Faster than hermitian quantum
  mechanics},\ }\href {https://doi.org/10.1103/PhysRevLett.98.040403}
  {\bibfield  {journal} {\bibinfo  {journal} {Phys. Rev. Lett.}\ }\textbf
  {\bibinfo {volume} {98}},\ \bibinfo {pages} {040403} (\bibinfo {year}
  {2007})}\BibitemShut {NoStop}%
\bibitem [{\citenamefont {Feng}\ \emph {et~al.}(2017)\citenamefont {Feng},
  \citenamefont {El-Ganainy},\ and\ \citenamefont {Ge}}]{Feng2017}%
  \BibitemOpen
  \bibfield  {author} {\bibinfo {author} {\bibfnamefont {L.}~\bibnamefont
  {Feng}}, \bibinfo {author} {\bibfnamefont {R.}~\bibnamefont {El-Ganainy}},\
  and\ \bibinfo {author} {\bibfnamefont {L.}~\bibnamefont {Ge}},\ }\bibfield
  {title} {\bibinfo {title} {Non-hermitian photonics based on parity--time
  symmetry},\ }\href {https://doi.org/10.1038/s41566-017-0031-1} {\bibfield
  {journal} {\bibinfo  {journal} {Nat. Photonics}\ }\textbf {\bibinfo {volume}
  {11}},\ \bibinfo {pages} {752} (\bibinfo {year} {2017})}\BibitemShut
  {NoStop}%
\bibitem [{\citenamefont {Motta}\ \emph {et~al.}(2019)\citenamefont {Motta},
  \citenamefont {Sun}, \citenamefont {Tan}, \citenamefont {O’Rourke},
  \citenamefont {Ye}, \citenamefont {Minnich}, \citenamefont {Brandão},\ and\
  \citenamefont {Chan}}]{Motta2019}%
  \BibitemOpen
  \bibfield  {author} {\bibinfo {author} {\bibfnamefont {M.}~\bibnamefont
  {Motta}}, \bibinfo {author} {\bibfnamefont {C.}~\bibnamefont {Sun}}, \bibinfo
  {author} {\bibfnamefont {A.~T.~K.}\ \bibnamefont {Tan}}, \bibinfo {author}
  {\bibfnamefont {M.~J.}\ \bibnamefont {O’Rourke}}, \bibinfo {author}
  {\bibfnamefont {E.}~\bibnamefont {Ye}}, \bibinfo {author} {\bibfnamefont
  {A.~J.}\ \bibnamefont {Minnich}}, \bibinfo {author} {\bibfnamefont {F.~G.
  S.~L.}\ \bibnamefont {Brandão}},\ and\ \bibinfo {author} {\bibfnamefont
  {G.~K.-L.}\ \bibnamefont {Chan}},\ }\bibfield  {title} {\bibinfo {title}
  {Determining eigenstates and thermal states on a quantum computer using
  quantum imaginary time evolution},\ }\href
  {https://doi.org/10.1038/s41567-019-0704-4} {\bibfield  {journal} {\bibinfo
  {journal} {Nature Physics}\ }\textbf {\bibinfo {volume} {16}},\ \bibinfo
  {pages} {205–210} (\bibinfo {year} {2019})}\BibitemShut {NoStop}%
\bibitem [{\citenamefont {del Campo}\ \emph {et~al.}(2013)\citenamefont {del
  Campo}, \citenamefont {Egusquiza}, \citenamefont {Plenio},\ and\
  \citenamefont {Huelga}}]{delcampo13}%
  \BibitemOpen
  \bibfield  {author} {\bibinfo {author} {\bibfnamefont {A.}~\bibnamefont {del
  Campo}}, \bibinfo {author} {\bibfnamefont {I.~L.}\ \bibnamefont {Egusquiza}},
  \bibinfo {author} {\bibfnamefont {M.~B.}\ \bibnamefont {Plenio}},\ and\
  \bibinfo {author} {\bibfnamefont {S.~F.}\ \bibnamefont {Huelga}},\ }\bibfield
   {title} {\bibinfo {title} {Quantum speed limits in open system dynamics},\
  }\href {https://doi.org/10.1103/PhysRevLett.110.050403} {\bibfield  {journal}
  {\bibinfo  {journal} {Phys. Rev. Lett.}\ }\textbf {\bibinfo {volume} {110}},\
  \bibinfo {pages} {050403} (\bibinfo {year} {2013})}\BibitemShut {NoStop}%
\bibitem [{\citenamefont {Hörnedal}\ \emph {et~al.}(2024)\citenamefont
  {Hörnedal}, \citenamefont {Prośniak}, \citenamefont {del Campo},\ and\
  \citenamefont {Chenu}}]{Hornedal2024}%
  \BibitemOpen
  \bibfield  {author} {\bibinfo {author} {\bibfnamefont {N.}~\bibnamefont
  {Hörnedal}}, \bibinfo {author} {\bibfnamefont {O.~A.}\ \bibnamefont
  {Prośniak}}, \bibinfo {author} {\bibfnamefont {A.}~\bibnamefont {del
  Campo}},\ and\ \bibinfo {author} {\bibfnamefont {A.}~\bibnamefont {Chenu}},\
  }\href {https://arxiv.org/abs/2405.13913} {\bibinfo {title} {A geometrical
  description of non-hermitian dynamics: speed limits in finite rank density
  operators}} (\bibinfo {year} {2024}),\ \Eprint
  {https://arxiv.org/abs/2405.13913} {arXiv:2405.13913 [quant-ph]} \BibitemShut
  {NoStop}%
\bibitem [{\citenamefont {Vacanti}\ \emph {et~al.}(2014)\citenamefont
  {Vacanti}, \citenamefont {Fazio}, \citenamefont {Montangero}, \citenamefont
  {Palma}, \citenamefont {Paternostro},\ and\ \citenamefont
  {Vedral}}]{Vacanti2014}%
  \BibitemOpen
  \bibfield  {author} {\bibinfo {author} {\bibfnamefont {G.}~\bibnamefont
  {Vacanti}}, \bibinfo {author} {\bibfnamefont {R.}~\bibnamefont {Fazio}},
  \bibinfo {author} {\bibfnamefont {S.}~\bibnamefont {Montangero}}, \bibinfo
  {author} {\bibfnamefont {G.~M.}\ \bibnamefont {Palma}}, \bibinfo {author}
  {\bibfnamefont {M.}~\bibnamefont {Paternostro}},\ and\ \bibinfo {author}
  {\bibfnamefont {V.}~\bibnamefont {Vedral}},\ }\bibfield  {title} {\bibinfo
  {title} {Transitionless quantum driving in open quantum systems},\ }\href
  {https://doi.org/10.1088/1367-2630/16/5/053017} {\bibfield  {journal}
  {\bibinfo  {journal} {New Journal of Physics}\ }\textbf {\bibinfo {volume}
  {16}},\ \bibinfo {pages} {053017} (\bibinfo {year} {2014})}\BibitemShut
  {NoStop}%
\bibitem [{\citenamefont {Dann}\ \emph {et~al.}(2019)\citenamefont {Dann},
  \citenamefont {Tobalina},\ and\ \citenamefont {Kosloff}}]{Dann19}%
  \BibitemOpen
  \bibfield  {author} {\bibinfo {author} {\bibfnamefont {R.}~\bibnamefont
  {Dann}}, \bibinfo {author} {\bibfnamefont {A.}~\bibnamefont {Tobalina}},\
  and\ \bibinfo {author} {\bibfnamefont {R.}~\bibnamefont {Kosloff}},\
  }\bibfield  {title} {\bibinfo {title} {Shortcut to equilibration of an open
  quantum system},\ }\href {https://doi.org/10.1103/PhysRevLett.122.250402}
  {\bibfield  {journal} {\bibinfo  {journal} {Phys. Rev. Lett.}\ }\textbf
  {\bibinfo {volume} {122}},\ \bibinfo {pages} {250402} (\bibinfo {year}
  {2019})}\BibitemShut {NoStop}%
\bibitem [{\citenamefont {Alipour}\ \emph {et~al.}(2020)\citenamefont
  {Alipour}, \citenamefont {Chenu}, \citenamefont {Rezakhani},\ and\
  \citenamefont {del Campo}}]{Alipour2020}%
  \BibitemOpen
  \bibfield  {author} {\bibinfo {author} {\bibfnamefont {S.}~\bibnamefont
  {Alipour}}, \bibinfo {author} {\bibfnamefont {A.}~\bibnamefont {Chenu}},
  \bibinfo {author} {\bibfnamefont {A.~T.}\ \bibnamefont {Rezakhani}},\ and\
  \bibinfo {author} {\bibfnamefont {A.}~\bibnamefont {del Campo}},\ }\bibfield
  {title} {\bibinfo {title} {Shortcuts to adiabaticity in driven open quantum
  systems: Balanced gain and loss and non-markovian evolution},\ }\href
  {https://doi.org/10.22331/q-2020-09-28-336} {\bibfield  {journal} {\bibinfo
  {journal} {Quantum}\ }\textbf {\bibinfo {volume} {4}},\ \bibinfo {pages}
  {336} (\bibinfo {year} {2020})}\BibitemShut {NoStop}%
\bibitem [{\citenamefont {Dupays}\ \emph {et~al.}(2020)\citenamefont {Dupays},
  \citenamefont {Egusquiza}, \citenamefont {del Campo},\ and\ \citenamefont
  {Chenu}}]{Dupays2020}%
  \BibitemOpen
  \bibfield  {author} {\bibinfo {author} {\bibfnamefont {L.}~\bibnamefont
  {Dupays}}, \bibinfo {author} {\bibfnamefont {I.~L.}\ \bibnamefont
  {Egusquiza}}, \bibinfo {author} {\bibfnamefont {A.}~\bibnamefont {del
  Campo}},\ and\ \bibinfo {author} {\bibfnamefont {A.}~\bibnamefont {Chenu}},\
  }\bibfield  {title} {\bibinfo {title} {Superadiabatic thermalization of a
  quantum oscillator by engineered dephasing},\ }\href
  {https://doi.org/10.1103/PhysRevResearch.2.033178} {\bibfield  {journal}
  {\bibinfo  {journal} {Phys. Rev. Res.}\ }\textbf {\bibinfo {volume} {2}},\
  \bibinfo {pages} {033178} (\bibinfo {year} {2020})}\BibitemShut {NoStop}%
\bibitem [{\citenamefont {Dupays}\ and\ \citenamefont
  {Chenu}(2021)}]{Dupays2021shortcutstosqueezed}%
  \BibitemOpen
  \bibfield  {author} {\bibinfo {author} {\bibfnamefont {L.}~\bibnamefont
  {Dupays}}\ and\ \bibinfo {author} {\bibfnamefont {A.}~\bibnamefont {Chenu}},\
  }\bibfield  {title} {\bibinfo {title} {Shortcuts to {S}queezed {T}hermal
  {S}tates},\ }\href {https://doi.org/10.22331/q-2021-05-01-449} {\bibfield
  {journal} {\bibinfo  {journal} {{Quantum}}\ }\textbf {\bibinfo {volume}
  {5}},\ \bibinfo {pages} {449} (\bibinfo {year} {2021})}\BibitemShut {NoStop}%
\bibitem [{\citenamefont {Yin}\ \emph {et~al.}(2022)\citenamefont {Yin},
  \citenamefont {Li}, \citenamefont {Allcock}, \citenamefont {Zheng},
  \citenamefont {Gu}, \citenamefont {Dai}, \citenamefont {Zhang},\ and\
  \citenamefont {An}}]{Yin2022}%
  \BibitemOpen
  \bibfield  {author} {\bibinfo {author} {\bibfnamefont {Z.}~\bibnamefont
  {Yin}}, \bibinfo {author} {\bibfnamefont {C.}~\bibnamefont {Li}}, \bibinfo
  {author} {\bibfnamefont {J.}~\bibnamefont {Allcock}}, \bibinfo {author}
  {\bibfnamefont {Y.}~\bibnamefont {Zheng}}, \bibinfo {author} {\bibfnamefont
  {X.}~\bibnamefont {Gu}}, \bibinfo {author} {\bibfnamefont {M.}~\bibnamefont
  {Dai}}, \bibinfo {author} {\bibfnamefont {S.}~\bibnamefont {Zhang}},\ and\
  \bibinfo {author} {\bibfnamefont {S.}~\bibnamefont {An}},\ }\bibfield
  {title} {\bibinfo {title} {Shortcuts to adiabaticity for open systems in
  circuit quantum electrodynamics},\ }\href
  {https://doi.org/10.1038/s41467-021-27900-6} {\bibfield  {journal} {\bibinfo
  {journal} {Nature Communications}\ }\textbf {\bibinfo {volume} {13}},\
  \bibinfo {pages} {188} (\bibinfo {year} {2022})}\BibitemShut {NoStop}%
\bibitem [{\citenamefont {Ib\'a\~nez}\ \emph {et~al.}(2011)\citenamefont
  {Ib\'a\~nez}, \citenamefont {Mart\'{\i}nez-Garaot}, \citenamefont {Chen},
  \citenamefont {Torrontegui},\ and\ \citenamefont {Muga}}]{Ibanez11}%
  \BibitemOpen
  \bibfield  {author} {\bibinfo {author} {\bibfnamefont {S.}~\bibnamefont
  {Ib\'a\~nez}}, \bibinfo {author} {\bibfnamefont {S.}~\bibnamefont
  {Mart\'{\i}nez-Garaot}}, \bibinfo {author} {\bibfnamefont {X.}~\bibnamefont
  {Chen}}, \bibinfo {author} {\bibfnamefont {E.}~\bibnamefont {Torrontegui}},\
  and\ \bibinfo {author} {\bibfnamefont {J.~G.}\ \bibnamefont {Muga}},\
  }\bibfield  {title} {\bibinfo {title} {Shortcuts to adiabaticity for
  non-hermitian systems},\ }\href {https://doi.org/10.1103/PhysRevA.84.023415}
  {\bibfield  {journal} {\bibinfo  {journal} {Phys. Rev. A}\ }\textbf {\bibinfo
  {volume} {84}},\ \bibinfo {pages} {023415} (\bibinfo {year}
  {2011})}\BibitemShut {NoStop}%
\bibitem [{\citenamefont {Torosov}\ \emph {et~al.}(2013)\citenamefont
  {Torosov}, \citenamefont {Della~Valle},\ and\ \citenamefont
  {Longhi}}]{Torosov13}%
  \BibitemOpen
  \bibfield  {author} {\bibinfo {author} {\bibfnamefont {B.~T.}\ \bibnamefont
  {Torosov}}, \bibinfo {author} {\bibfnamefont {G.}~\bibnamefont
  {Della~Valle}},\ and\ \bibinfo {author} {\bibfnamefont {S.}~\bibnamefont
  {Longhi}},\ }\bibfield  {title} {\bibinfo {title} {Non-hermitian shortcut to
  adiabaticity},\ }\href {https://doi.org/10.1103/PhysRevA.87.052502}
  {\bibfield  {journal} {\bibinfo  {journal} {Phys. Rev. A}\ }\textbf {\bibinfo
  {volume} {87}},\ \bibinfo {pages} {052502} (\bibinfo {year}
  {2013})}\BibitemShut {NoStop}%
\bibitem [{\citenamefont {Hatano}\ and\ \citenamefont
  {Nelson}(1997)}]{hatanonelson1}%
  \BibitemOpen
  \bibfield  {author} {\bibinfo {author} {\bibfnamefont {N.}~\bibnamefont
  {Hatano}}\ and\ \bibinfo {author} {\bibfnamefont {D.~R.}\ \bibnamefont
  {Nelson}},\ }\bibfield  {title} {\bibinfo {title} {Vortex pinning and
  non-hermitian quantum mechanics},\ }\href
  {https://doi.org/10.1103/PhysRevB.56.8651} {\bibfield  {journal} {\bibinfo
  {journal} {Phys. Rev. B}\ }\textbf {\bibinfo {volume} {56}},\ \bibinfo
  {pages} {8651} (\bibinfo {year} {1997})}\BibitemShut {NoStop}%
\bibitem [{\citenamefont {Hatano}\ and\ \citenamefont
  {Nelson}(1996)}]{hatanonelson2}%
  \BibitemOpen
  \bibfield  {author} {\bibinfo {author} {\bibfnamefont {N.}~\bibnamefont
  {Hatano}}\ and\ \bibinfo {author} {\bibfnamefont {D.~R.}\ \bibnamefont
  {Nelson}},\ }\bibfield  {title} {\bibinfo {title} {Localization transitions
  in non-hermitian quantum mechanics},\ }\href
  {https://doi.org/10.1103/PhysRevLett.77.570} {\bibfield  {journal} {\bibinfo
  {journal} {Phys. Rev. Lett.}\ }\textbf {\bibinfo {volume} {77}},\ \bibinfo
  {pages} {570} (\bibinfo {year} {1996})}\BibitemShut {NoStop}%
\bibitem [{\citenamefont {Shen}\ \emph {et~al.}()\citenamefont {Shen},
  \citenamefont {Chen}, \citenamefont {Yang},\ and\ \citenamefont
  {Lee}}]{shen2023}%
  \BibitemOpen
  \bibfield  {author} {\bibinfo {author} {\bibfnamefont {R.}~\bibnamefont
  {Shen}}, \bibinfo {author} {\bibfnamefont {T.}~\bibnamefont {Chen}}, \bibinfo
  {author} {\bibfnamefont {B.}~\bibnamefont {Yang}},\ and\ \bibinfo {author}
  {\bibfnamefont {C.~H.}\ \bibnamefont {Lee}},\ }\bibfield  {title} {\bibinfo
  {title} {Observation of the non-hermitian skin effect and fermi skin on a
  digital quantum computer},\ }\bibinfo {note} {ar{X}iv:2311.10143}\BibitemShut
  {NoStop}%
\bibitem [{\citenamefont {Ren}\ \emph {et~al.}(2022)\citenamefont {Ren},
  \citenamefont {Liu}, \citenamefont {Zhao}, \citenamefont {He}, \citenamefont
  {Pak}, \citenamefont {Li},\ and\ \citenamefont {Jo}}]{ren2022}%
  \BibitemOpen
  \bibfield  {author} {\bibinfo {author} {\bibfnamefont {Z.}~\bibnamefont
  {Ren}}, \bibinfo {author} {\bibfnamefont {D.}~\bibnamefont {Liu}}, \bibinfo
  {author} {\bibfnamefont {E.}~\bibnamefont {Zhao}}, \bibinfo {author}
  {\bibfnamefont {C.}~\bibnamefont {He}}, \bibinfo {author} {\bibfnamefont
  {K.~K.}\ \bibnamefont {Pak}}, \bibinfo {author} {\bibfnamefont
  {J.}~\bibnamefont {Li}},\ and\ \bibinfo {author} {\bibfnamefont {G.-B.}\
  \bibnamefont {Jo}},\ }\bibfield  {title} {\bibinfo {title} {Chiral control of
  quantum states in non-hermitian spin–orbit-coupled fermions},\ }\href
  {https://doi.org/10.1038/s41567-021-01491-x} {\bibfield  {journal} {\bibinfo
  {journal} {Nat. Phys.}\ }\textbf {\bibinfo {volume} {18}},\ \bibinfo {pages}
  {385} (\bibinfo {year} {2022})}\BibitemShut {NoStop}%
\bibitem [{\citenamefont {Gou}\ \emph {et~al.}(2020)\citenamefont {Gou},
  \citenamefont {Chen}, \citenamefont {Xie}, \citenamefont {Xiao},
  \citenamefont {Deng}, \citenamefont {Gadway}, \citenamefont {Yi},\ and\
  \citenamefont {Yan}}]{guo2020}%
  \BibitemOpen
  \bibfield  {author} {\bibinfo {author} {\bibfnamefont {W.}~\bibnamefont
  {Gou}}, \bibinfo {author} {\bibfnamefont {T.}~\bibnamefont {Chen}}, \bibinfo
  {author} {\bibfnamefont {D.}~\bibnamefont {Xie}}, \bibinfo {author}
  {\bibfnamefont {T.}~\bibnamefont {Xiao}}, \bibinfo {author} {\bibfnamefont
  {T.-S.}\ \bibnamefont {Deng}}, \bibinfo {author} {\bibfnamefont
  {B.}~\bibnamefont {Gadway}}, \bibinfo {author} {\bibfnamefont
  {W.}~\bibnamefont {Yi}},\ and\ \bibinfo {author} {\bibfnamefont
  {B.}~\bibnamefont {Yan}},\ }\bibfield  {title} {\bibinfo {title} {Tunable
  nonreciprocal quantum transport through a dissipative aharonov-bohm ring in
  ultracold atoms},\ }\href {https://doi.org/10.1103/PhysRevLett.124.070402}
  {\bibfield  {journal} {\bibinfo  {journal} {Phys. Rev. Lett.}\ }\textbf
  {\bibinfo {volume} {124}},\ \bibinfo {pages} {070402} (\bibinfo {year}
  {2020})}\BibitemShut {NoStop}%
\bibitem [{\citenamefont {Liang}\ \emph {et~al.}(2022)\citenamefont {Liang},
  \citenamefont {Xie}, \citenamefont {Dong}, \citenamefont {Li}, \citenamefont
  {Li}, \citenamefont {Gadway}, \citenamefont {Yi},\ and\ \citenamefont
  {Yan}}]{liang2022}%
  \BibitemOpen
  \bibfield  {author} {\bibinfo {author} {\bibfnamefont {Q.}~\bibnamefont
  {Liang}}, \bibinfo {author} {\bibfnamefont {D.}~\bibnamefont {Xie}}, \bibinfo
  {author} {\bibfnamefont {Z.}~\bibnamefont {Dong}}, \bibinfo {author}
  {\bibfnamefont {H.}~\bibnamefont {Li}}, \bibinfo {author} {\bibfnamefont
  {H.}~\bibnamefont {Li}}, \bibinfo {author} {\bibfnamefont {B.}~\bibnamefont
  {Gadway}}, \bibinfo {author} {\bibfnamefont {W.}~\bibnamefont {Yi}},\ and\
  \bibinfo {author} {\bibfnamefont {B.}~\bibnamefont {Yan}},\ }\bibfield
  {title} {\bibinfo {title} {Dynamic signatures of non-hermitian skin effect
  and topology in ultracold atoms},\ }\href
  {https://doi.org/10.1103/PhysRevLett.129.070401} {\bibfield  {journal}
  {\bibinfo  {journal} {Phys. Rev. Lett.}\ }\textbf {\bibinfo {volume} {129}},\
  \bibinfo {pages} {070401} (\bibinfo {year} {2022})}\BibitemShut {NoStop}%
\bibitem [{\citenamefont {Giamarchi}(2004)}]{giamarchi}%
  \BibitemOpen
  \bibfield  {author} {\bibinfo {author} {\bibfnamefont {T.}~\bibnamefont
  {Giamarchi}},\ }\href@noop {} {\emph {\bibinfo {title} {Quantum Physics in
  One Dimension}}}\ (\bibinfo  {publisher} {Oxford University Press},\ \bibinfo
  {address} {Oxford},\ \bibinfo {year} {2004})\BibitemShut {NoStop}%
\bibitem [{\citenamefont {Cazalilla}(2004)}]{cazalillaboson}%
  \BibitemOpen
  \bibfield  {author} {\bibinfo {author} {\bibfnamefont {M.~A.}\ \bibnamefont
  {Cazalilla}},\ }\bibfield  {title} {\bibinfo {title} {Bosonizing
  one-dimensional cold atomic gases},\ }\href
  {https://doi.org/10.1088/0953-4075/37/7/051} {\bibfield  {journal} {\bibinfo
  {journal} {J. Phys. B}\ }\textbf {\bibinfo {volume} {37}},\ \bibinfo {pages}
  {S1} (\bibinfo {year} {2004})}\BibitemShut {NoStop}%
\bibitem [{\citenamefont {Gogolin}\ \emph {et~al.}(1998)\citenamefont
  {Gogolin}, \citenamefont {Nersesyan},\ and\ \citenamefont
  {Tsvelik}}]{nersesyan}%
  \BibitemOpen
  \bibfield  {author} {\bibinfo {author} {\bibfnamefont {A.~O.}\ \bibnamefont
  {Gogolin}}, \bibinfo {author} {\bibfnamefont {A.~A.}\ \bibnamefont
  {Nersesyan}},\ and\ \bibinfo {author} {\bibfnamefont {A.~M.}\ \bibnamefont
  {Tsvelik}},\ }\href@noop {} {\emph {\bibinfo {title} {Bosonization and
  Strongly Correlated Systems}}}\ (\bibinfo  {publisher} {Cambridge University
  Press},\ \bibinfo {address} {Cambridge},\ \bibinfo {year} {1998})\BibitemShut
  {NoStop}%
\bibitem [{\citenamefont {Hofstetter}\ \emph {et~al.}(2004)\citenamefont
  {Hofstetter}, \citenamefont {Affleck}, \citenamefont {Nelson},\ and\
  \citenamefont {Schollwöck}}]{Hofstetter2004}%
  \BibitemOpen
  \bibfield  {author} {\bibinfo {author} {\bibfnamefont {W.}~\bibnamefont
  {Hofstetter}}, \bibinfo {author} {\bibfnamefont {I.}~\bibnamefont {Affleck}},
  \bibinfo {author} {\bibfnamefont {D.}~\bibnamefont {Nelson}},\ and\ \bibinfo
  {author} {\bibfnamefont {U.}~\bibnamefont {Schollwöck}},\ }\bibfield
  {title} {\bibinfo {title} {Non-hermitian luttinger liquids and vortex
  physics},\ }\href {https://doi.org/10.1209/epl/i2003-10204-2} {\bibfield
  {journal} {\bibinfo  {journal} {EPL}\ }\textbf {\bibinfo {volume} {66}},\
  \bibinfo {pages} {178} (\bibinfo {year} {2004})}\BibitemShut {NoStop}%
\bibitem [{\citenamefont {Yamamoto}\ \emph {et~al.}(2022)\citenamefont
  {Yamamoto}, \citenamefont {Nakagawa}, \citenamefont {Tezuka}, \citenamefont
  {Ueda},\ and\ \citenamefont {Kawakami}}]{yamamoto}%
  \BibitemOpen
  \bibfield  {author} {\bibinfo {author} {\bibfnamefont {K.}~\bibnamefont
  {Yamamoto}}, \bibinfo {author} {\bibfnamefont {M.}~\bibnamefont {Nakagawa}},
  \bibinfo {author} {\bibfnamefont {M.}~\bibnamefont {Tezuka}}, \bibinfo
  {author} {\bibfnamefont {M.}~\bibnamefont {Ueda}},\ and\ \bibinfo {author}
  {\bibfnamefont {N.}~\bibnamefont {Kawakami}},\ }\bibfield  {title} {\bibinfo
  {title} {Universal properties of dissipative tomonaga-luttinger liquids: Case
  study of a non-hermitian xxz spin chain},\ }\href
  {https://doi.org/10.1103/PhysRevB.105.205125} {\bibfield  {journal} {\bibinfo
   {journal} {Phys. Rev. B}\ }\textbf {\bibinfo {volume} {105}},\ \bibinfo
  {pages} {205125} (\bibinfo {year} {2022})}\BibitemShut {NoStop}%
\bibitem [{\citenamefont {D\'ora}\ \emph {et~al.}(2023)\citenamefont {D\'ora},
  \citenamefont {Werner},\ and\ \citenamefont {Moca}}]{Dora2023PRB}%
  \BibitemOpen
  \bibfield  {author} {\bibinfo {author} {\bibfnamefont {B.}~\bibnamefont
  {D\'ora}}, \bibinfo {author} {\bibfnamefont {M.~A.}\ \bibnamefont {Werner}},\
  and\ \bibinfo {author} {\bibfnamefont {C.~P.}\ \bibnamefont {Moca}},\
  }\bibfield  {title} {\bibinfo {title} {Quantum quench dynamics in the
  luttinger liquid phase of the hatano-nelson model},\ }\href
  {https://doi.org/10.1103/PhysRevB.108.035104} {\bibfield  {journal} {\bibinfo
   {journal} {Phys. Rev. B}\ }\textbf {\bibinfo {volume} {108}},\ \bibinfo
  {pages} {035104} (\bibinfo {year} {2023})}\BibitemShut {NoStop}%
\bibitem [{\citenamefont {Schomerus}\ and\ \citenamefont
  {Wiersig}(2014)}]{schomerus}%
  \BibitemOpen
  \bibfield  {author} {\bibinfo {author} {\bibfnamefont {H.}~\bibnamefont
  {Schomerus}}\ and\ \bibinfo {author} {\bibfnamefont {J.}~\bibnamefont
  {Wiersig}},\ }\bibfield  {title} {\bibinfo {title} {Non-hermitian-transport
  effects in coupled-resonator optical waveguides},\ }\href
  {https://doi.org/10.1103/PhysRevA.90.053819} {\bibfield  {journal} {\bibinfo
  {journal} {Phys. Rev. A}\ }\textbf {\bibinfo {volume} {90}},\ \bibinfo
  {pages} {053819} (\bibinfo {year} {2014})}\BibitemShut {NoStop}%
\bibitem [{\citenamefont {Dóra}\ and\ \citenamefont
  {Moca}(2023)}]{Dora2023work}%
  \BibitemOpen
  \bibfield  {author} {\bibinfo {author} {\bibfnamefont {B.}~\bibnamefont
  {Dóra}}\ and\ \bibinfo {author} {\bibfnamefont {C.~P.}\ \bibnamefont
  {Moca}},\ }\bibfield  {title} {\bibinfo {title} {Work statistics and
  generalized loschmidt echo for the hatano-nelson model},\ }\href@noop {} {\
  (\bibinfo {year} {2023})},\ \Eprint {https://arxiv.org/abs/2310.19310}
  {arXiv:2310.19310} \BibitemShut {NoStop}%
\bibitem [{\citenamefont {Gradshteyn}\ and\ \citenamefont
  {Ryzhik}(2007)}]{gradstein}%
  \BibitemOpen
  \bibfield  {author} {\bibinfo {author} {\bibfnamefont {I.}~\bibnamefont
  {Gradshteyn}}\ and\ \bibinfo {author} {\bibfnamefont {I.}~\bibnamefont
  {Ryzhik}},\ }\href@noop {} {\emph {\bibinfo {title} {Table of Integrals,
  Series, and Products}}}\ (\bibinfo  {publisher} {Academic Press},\ \bibinfo
  {address} {New York},\ \bibinfo {year} {2007})\BibitemShut {NoStop}%
\bibitem [{\citenamefont {Misra}\ and\ \citenamefont
  {Sudarshan}(1977)}]{misra}%
  \BibitemOpen
  \bibfield  {author} {\bibinfo {author} {\bibfnamefont {B.}~\bibnamefont
  {Misra}}\ and\ \bibinfo {author} {\bibfnamefont {E.~C.~G.}\ \bibnamefont
  {Sudarshan}},\ }\bibfield  {title} {\bibinfo {title} {The zeno’s paradox in
  quantum theory},\ }\href {https://doi.org/10.1063/1.523304} {\bibfield
  {journal} {\bibinfo  {journal} {Journal of Mathematical Physics}\ }\textbf
  {\bibinfo {volume} {18}},\ \bibinfo {pages} {756} (\bibinfo {year}
  {1977})}\BibitemShut {NoStop}%
\bibitem [{\citenamefont {Jaramillo}\ \emph {et~al.}(2016)\citenamefont
  {Jaramillo}, \citenamefont {Beau},\ and\ \citenamefont {del
  Campo}}]{Jaramillo_2016}%
  \BibitemOpen
  \bibfield  {author} {\bibinfo {author} {\bibfnamefont {J.}~\bibnamefont
  {Jaramillo}}, \bibinfo {author} {\bibfnamefont {M.}~\bibnamefont {Beau}},\
  and\ \bibinfo {author} {\bibfnamefont {A.}~\bibnamefont {del Campo}},\
  }\bibfield  {title} {\bibinfo {title} {Quantum supremacy of many-particle
  thermal machines},\ }\href {https://doi.org/10.1088/1367-2630/18/7/075019}
  {\bibfield  {journal} {\bibinfo  {journal} {New J. Phys.}\ }\textbf {\bibinfo
  {volume} {18}},\ \bibinfo {pages} {075019} (\bibinfo {year}
  {2016})}\BibitemShut {NoStop}%
\bibitem [{\citenamefont {Dupays}\ \emph {et~al.}(2024)\citenamefont {Dupays},
  \citenamefont {Dóra},\ and\ \citenamefont {del Campo}}]{Dupays2024exact}%
  \BibitemOpen
  \bibfield  {author} {\bibinfo {author} {\bibfnamefont {L.}~\bibnamefont
  {Dupays}}, \bibinfo {author} {\bibfnamefont {B.}~\bibnamefont {Dóra}},\ and\
  \bibinfo {author} {\bibfnamefont {A.}~\bibnamefont {del Campo}},\ }\bibfield
  {title} {\bibinfo {title} {Exact dynamics and shortcuts to adiabaticity in
  the tomonaga-luttinger liquid},\ }\href@noop {} {\  (\bibinfo {year}
  {2024})},\ \Eprint {https://arxiv.org/abs/2401.17884} {arXiv:2401.17884}
  \BibitemShut {NoStop}%
\bibitem [{\citenamefont {Dupays}\ and\ \citenamefont {del
  Campo}(2024)}]{Dupays2024transitionless}%
  \BibitemOpen
  \bibfield  {author} {\bibinfo {author} {\bibfnamefont {L.}~\bibnamefont
  {Dupays}}\ and\ \bibinfo {author} {\bibfnamefont {A.}~\bibnamefont {del
  Campo}},\ }\bibfield  {title} {\bibinfo {title} {Transitionless quantum
  driving of the tomonaga-luttinger liquid},\ }\href@noop {} {\  (\bibinfo
  {year} {2024})},\ \Eprint {https://arxiv.org/abs/2404.19013}
  {arXiv:2404.19013} \BibitemShut {NoStop}%
\bibitem [{\citenamefont {Alsallom}\ \emph {et~al.}(2022)\citenamefont
  {Alsallom}, \citenamefont {Herviou}, \citenamefont {Yazyev},\ and\
  \citenamefont {Brzezi\ifmmode~\acute{n}\else \'{n}\fi{}ska}}]{alsallom}%
  \BibitemOpen
  \bibfield  {author} {\bibinfo {author} {\bibfnamefont {F.}~\bibnamefont
  {Alsallom}}, \bibinfo {author} {\bibfnamefont {L.}~\bibnamefont {Herviou}},
  \bibinfo {author} {\bibfnamefont {O.~V.}\ \bibnamefont {Yazyev}},\ and\
  \bibinfo {author} {\bibfnamefont {M.}~\bibnamefont
  {Brzezi\ifmmode~\acute{n}\else \'{n}\fi{}ska}},\ }\bibfield  {title}
  {\bibinfo {title} {Fate of the non-hermitian skin effect in many-body
  fermionic systems},\ }\href
  {https://doi.org/10.1103/PhysRevResearch.4.033122} {\bibfield  {journal}
  {\bibinfo  {journal} {Phys. Rev. Res.}\ }\textbf {\bibinfo {volume} {4}},\
  \bibinfo {pages} {033122} (\bibinfo {year} {2022})}\BibitemShut {NoStop}%
\bibitem [{\citenamefont {Silva}(2008)}]{silva}%
  \BibitemOpen
  \bibfield  {author} {\bibinfo {author} {\bibfnamefont {A.}~\bibnamefont
  {Silva}},\ }\bibfield  {title} {\bibinfo {title} {Statistics of the work done
  on a quantum critical system by quenching a control parameter},\ }\href
  {https://doi.org/10.1103/PhysRevLett.101.120603} {\bibfield  {journal}
  {\bibinfo  {journal} {Phys. Rev. Lett.}\ }\textbf {\bibinfo {volume} {101}},\
  \bibinfo {pages} {120603} (\bibinfo {year} {2008})}\BibitemShut {NoStop}%
\bibitem [{\citenamefont {Campisi}\ \emph {et~al.}(2011)\citenamefont
  {Campisi}, \citenamefont {H\"anggi},\ and\ \citenamefont
  {Talkner}}]{rmptalkner}%
  \BibitemOpen
  \bibfield  {author} {\bibinfo {author} {\bibfnamefont {M.}~\bibnamefont
  {Campisi}}, \bibinfo {author} {\bibfnamefont {P.}~\bibnamefont {H\"anggi}},\
  and\ \bibinfo {author} {\bibfnamefont {P.}~\bibnamefont {Talkner}},\
  }\bibfield  {title} {\bibinfo {title} {\textit{Colloquium} : Quantum
  fluctuation relations: Foundations and applications},\ }\href
  {https://doi.org/10.1103/RevModPhys.83.771} {\bibfield  {journal} {\bibinfo
  {journal} {Rev. Mod. Phys.}\ }\textbf {\bibinfo {volume} {83}},\ \bibinfo
  {pages} {771} (\bibinfo {year} {2011})}\BibitemShut {NoStop}%
\bibitem [{\citenamefont {Goussev}\ \emph {et~al.}(2012)\citenamefont
  {Goussev}, \citenamefont {Jalabert}, \citenamefont {Pastawski},\ and\
  \citenamefont {Wisniacki}}]{goussev}%
  \BibitemOpen
  \bibfield  {author} {\bibinfo {author} {\bibfnamefont {A.}~\bibnamefont
  {Goussev}}, \bibinfo {author} {\bibfnamefont {R.~A.}\ \bibnamefont
  {Jalabert}}, \bibinfo {author} {\bibfnamefont {H.~M.}\ \bibnamefont
  {Pastawski}},\ and\ \bibinfo {author} {\bibfnamefont {D.~A.}\ \bibnamefont
  {Wisniacki}},\ }\bibfield  {title} {\bibinfo {title} {Loschmidt echo},\
  }\href {https://doi.org/10.4249/scholarpedia.11687} {\bibfield  {journal}
  {\bibinfo  {journal} {Scholarpedia}\ }\textbf {\bibinfo {volume} {7}},\
  \bibinfo {pages} {11687} (\bibinfo {year} {2012})}\BibitemShut {NoStop}%
\bibitem [{\citenamefont {D\'ora}\ \emph {et~al.}(2013)\citenamefont {D\'ora},
  \citenamefont {Pollmann}, \citenamefont {Fort\'agh},\ and\ \citenamefont
  {Zar\'and}}]{doraLE}%
  \BibitemOpen
  \bibfield  {author} {\bibinfo {author} {\bibfnamefont {B.}~\bibnamefont
  {D\'ora}}, \bibinfo {author} {\bibfnamefont {F.}~\bibnamefont {Pollmann}},
  \bibinfo {author} {\bibfnamefont {J.}~\bibnamefont {Fort\'agh}},\ and\
  \bibinfo {author} {\bibfnamefont {G.}~\bibnamefont {Zar\'and}},\ }\bibfield
  {title} {\bibinfo {title} {Loschmidt echo and the many-body orthogonality
  catastrophe in a qubit-coupled luttinger liquid},\ }\href
  {https://doi.org/10.1103/PhysRevLett.111.046402} {\bibfield  {journal}
  {\bibinfo  {journal} {Phys. Rev. Lett.}\ }\textbf {\bibinfo {volume} {111}},\
  \bibinfo {pages} {046402} (\bibinfo {year} {2013})}\BibitemShut {NoStop}%
\bibitem [{\citenamefont {Tonielli}\ \emph {et~al.}(2020)\citenamefont
  {Tonielli}, \citenamefont {Chakraborty}, \citenamefont {Grusdt},\ and\
  \citenamefont {Marino}}]{Tonielli2020}%
  \BibitemOpen
  \bibfield  {author} {\bibinfo {author} {\bibfnamefont {F.}~\bibnamefont
  {Tonielli}}, \bibinfo {author} {\bibfnamefont {N.}~\bibnamefont
  {Chakraborty}}, \bibinfo {author} {\bibfnamefont {F.}~\bibnamefont
  {Grusdt}},\ and\ \bibinfo {author} {\bibfnamefont {J.}~\bibnamefont
  {Marino}},\ }\bibfield  {title} {\bibinfo {title} {Ramsey interferometry of
  non-hermitian quantum impurities},\ }\href
  {https://doi.org/10.1103/PhysRevResearch.2.032003} {\bibfield  {journal}
  {\bibinfo  {journal} {Phys. Rev. Res.}\ }\textbf {\bibinfo {volume} {2}},\
  \bibinfo {pages} {032003} (\bibinfo {year} {2020})}\BibitemShut {NoStop}%
\bibitem [{\citenamefont {Ness}\ \emph {et~al.}(2021)\citenamefont {Ness},
  \citenamefont {Lam}, \citenamefont {Alt}, \citenamefont {Meschede},
  \citenamefont {Sagi},\ and\ \citenamefont {Alberti}}]{Ness21}%
  \BibitemOpen
  \bibfield  {author} {\bibinfo {author} {\bibfnamefont {G.}~\bibnamefont
  {Ness}}, \bibinfo {author} {\bibfnamefont {M.~R.}\ \bibnamefont {Lam}},
  \bibinfo {author} {\bibfnamefont {W.}~\bibnamefont {Alt}}, \bibinfo {author}
  {\bibfnamefont {D.}~\bibnamefont {Meschede}}, \bibinfo {author}
  {\bibfnamefont {Y.}~\bibnamefont {Sagi}},\ and\ \bibinfo {author}
  {\bibfnamefont {A.}~\bibnamefont {Alberti}},\ }\bibfield  {title} {\bibinfo
  {title} {Observing crossover between quantum speed limits},\ }\href
  {https://doi.org/10.1126/sciadv.abj9119} {\bibfield  {journal} {\bibinfo
  {journal} {Science Advances}\ }\textbf {\bibinfo {volume} {7}},\ \bibinfo
  {pages} {eabj9119} (\bibinfo {year} {2021})}\BibitemShut {NoStop}%
\end{thebibliography}%

\title{Supplemental Material}
\maketitle
\onecolumngrid
\section{Fourier transform of the bosonized Hatano-Nelson Hamiltonian}
The bosonized Hatano-Nelson model is given by ({\color{RubineRed}$2$}) and the Fourier decomposition of the fields for open boundary conditions is given by ({\color{RubineRed}$3$a}) and ({\color{RubineRed}$3$b}), that can be further written 
\begin{align}
\Pi(x)&=\sqrt{\frac{2}{L}}\sum_{q>0}\sin qx\ \Pi_{q},\\
\phi(x)&=\sqrt{\frac{2}{L}}\sum_{q>0}\sin qx\ \phi_{q},
\end{align}
with $\Pi_{q}=-\sqrt{\frac{q}{2\pi K}}(b_{q}+b^{\dagger}_{q})$ and $\phi_{q}=i\sqrt{\frac{\pi K}{2q}}(b_{q}-b^{\dagger}_{q})$. Finally, the Hamiltonian is written as
\begin{align}
H(t)&=H_{\rm LL}-\sum_{q>0}\frac{2ih(t)vK\pi}{2\pi}\sqrt{\frac{2}{L}}\frac{(1-\cos qL)}{q}\Pi_{q}-\frac{vK h^{2}(t)L}{2\pi}\\
&=H_{\rm LL}+\sum_{q>0}ig_{q}(t)(b_{q}+b^{\dagger}_{q})-\frac{vK h^{2}(t)L}{2\pi}\\
&=\sum_{q>0}\frac{vK\pi}{2}\left(\Pi_{q}-i\frac{h(t)}{\pi}\sqrt{\frac{2}{L}}\frac{1-\cos qL}{q}\right)^{2}+\sum_{q>0}\frac{vq^{2}}{2\pi K}\phi^{2}_{q}+C_{0}(t),
\end{align}
with $C_{0}(t)=\frac{vK\pi}{2}F^{2}_{q}-\frac{vK h^{2}(t)L}{2\pi}$, $\sqrt{\frac{(vq)(vK\pi)}{2}}F_{q}=g_{q}(t)=\sqrt{\frac{K}{\pi L}}h(t)v\frac{(1-\cos qL)}{\sqrt{q}}$ and the Luttinger liquid (LL) Hamiltonian 
\begin{align}
H_{\rm LL}&=\sum_{q>0}\frac{v}{2\pi}\left[K\pi^{2}\Pi^{2}_{q}+\frac{ q^{2}}{K}\phi^{2}_{q}\right].\label{LL}
\end{align}
\section{Exact dynamics of the quenched Hatano-Nelson model}
As an alternative demonstration to the one provided in the main text, one can determine the dynamics in the $\Pi_{q}$ and $\phi_{q}$ representation. Let us consider the non-unitary dynamical map $V|0\rangle$, where $|0\rangle$ is the ground state and $VV^{-1}=\mathbbm{1}$. The initial Hamiltonian is of the form of a Luttinger liquid (LL) ({\color{RubineRed}$24$}) where the operators $\phi_{q}$ and $\Pi_{q}$ verify the canonical commutation relations $[\Pi_{q},\phi_{q'}]=i\delta_{qq'}$. One can build the dynamics by introducing the transformation  
\begin{align}
V=e^{-\sum_{q>0}\frac{K\pi}{vq^{2}}\dot{f}\Pi_{q}}e^{\sum_{q>0}f\phi_{q}}e^{\Phi},
\end{align}
so that 
\begin{align}
i\frac{d}{dt}(V|0\rangle)&=\left[VH_{0}V^{-1}+i\frac{dV}{dt}V^{-1}\right]V|0\rangle,\\
i\frac{d}{dt}|\Psi(t)\rangle&=\sum_{q>0}\left[\frac{vK\pi}{2}(\Pi_{q}-iF_{q})^{2}+\frac{vq^{2}}{2\pi K}\phi^{2}_{q}+C_{0}(t)\right]|\Psi(t)\rangle,
\end{align}
with
\begin{align}
F_{q}&=\frac{\ddot{f}}{(vq)^{2}}+f,\\
\\
i\frac{d}{dt}\Phi&=\frac{vK\pi}{2} f^{2}-\frac{1}{2}\frac{K\pi}{v q^{2}}(\dot{f})^{2}-\frac{vK h^{2}(t)L}{2\pi}.
\end{align}
Furthermore, we can now use the mode expansion $\Pi_{q}=-\sqrt{\frac{q}{2\pi K}}[b_{q}+b^{\dagger}_{q}]$ and $\phi_{q}=i\sqrt{\frac{\pi K}{2q}}[b_{q}-b^{\dagger}_{q}]$. As a consequence, we obtain 
\begin{align}
V&=e^{-\sum_{q>0}\frac{K\pi}{vq^{2}}\dot{f}\Pi_{q}}e^{\sum_{q>0}f\phi_{q}}e^{\Phi}\\
&=e^{-\sum_{q>0}\frac{K\pi}{vq^{2}}\dot{f}\Pi_{q}+\sum_{q>0}f\phi_{q}}e^{-\sum_{q>0}\frac{iK\pi}{2v q^{2}}\dot{f}f}e^{\Phi}\\
&=e^{\sum_{q>0}\sqrt{\frac{\pi K}{2q}}[\frac{\dot{f}}{vq}-if]b^{\dagger}_{q}+\sqrt{\frac{\pi K}{2q}}[\frac{\dot{f}}{vq}+if]b_{q}}e^{-\sum_{q>0}\frac{iK\pi}{2v q^{2}}\dot{f}f}e^{\Phi}\\
&=e^{\sum_{q>0}\sqrt{\frac{\pi K}{2q}}[\frac{\dot{f}}{vq}-if]b^{\dagger}_{q}}e^{\sqrt{\frac{\pi K}{2q}}[\frac{\dot{f}}{vq}+if]b_{q}}e^{\sum_{q>0}\frac{\pi K}{4q}[(\frac{\dot{f}}{vq})^{2}+f^{2}]}e^{-\sum_{q>0}\frac{iK\pi}{2v q^{2}}\dot{f}f}e^{\Phi}.
\end{align}
If we denote $\alpha_{q}=\sqrt{\frac{\pi K}{2q}}[\frac{\dot{f}}{vq}-if]$, one can verify that
\begin{align}
\dot{\alpha}_{q}+ivq\alpha_{q}=\sqrt{\frac{v\pi K (vq)}{2}}F_{q}=g_{q}(t).
\end{align}
Finally, the solution to this differential equation is given by the expression of the main text ({\color{RubineRed}$8$}).
\section{Mean energy in the Hatano-Nelson model with a linear quench}
\begin{figure}
    \centering
    \includegraphics[width=0.6\textwidth]{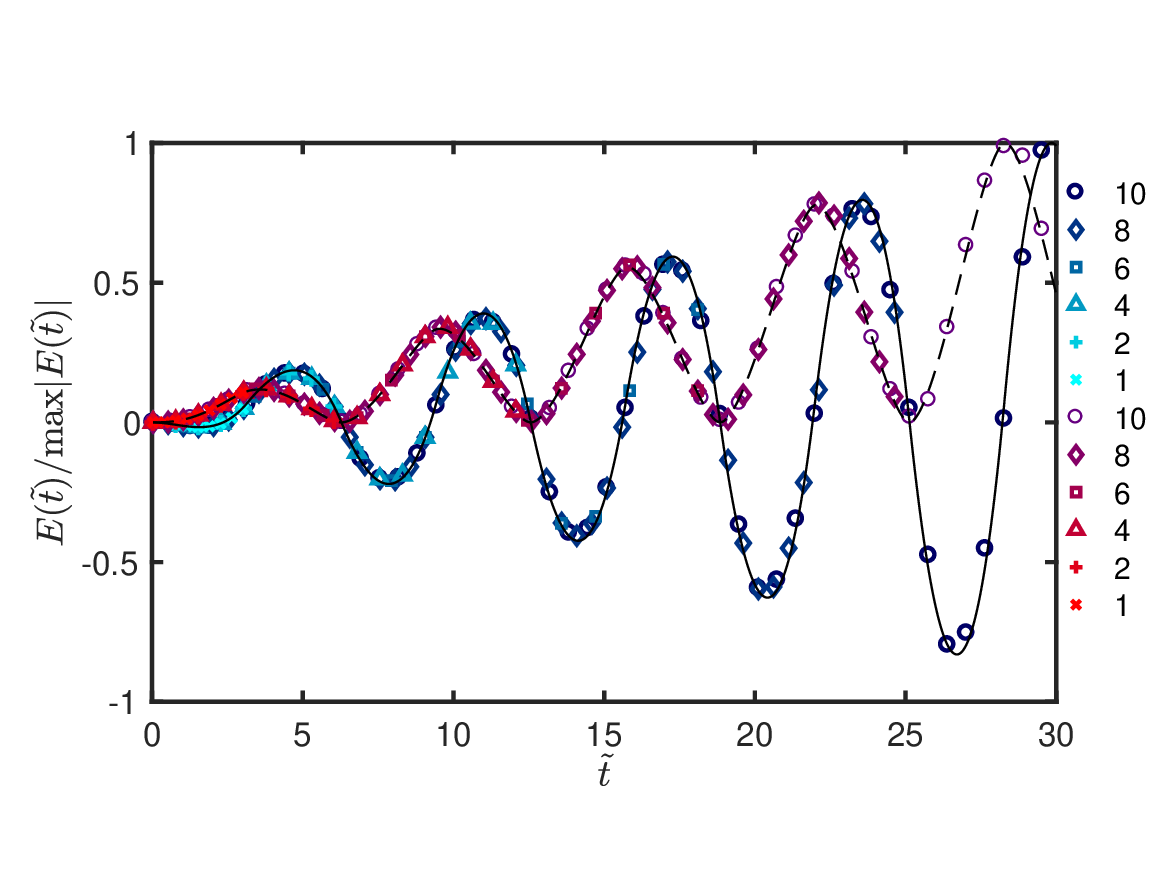}
    \caption{Evolution of the energy during the linear quench. The numerical simulation is plotted for different quench durations $\tau_{n}=nL/v$ with $n$ indicated in the legend and evaluated at the dimensionless time $\tilde{t}$. The imaginary part is plotted in nuances of red, and the real part in nuances of blue. Comparison of the bosonized real part in plain black line and the bosonized imaginary part in dashed black line. }
    \label{fig:transit}
\end{figure}

Let us now compute the mean energy for the quenched Hatano-Nelson model
\begin{align}
\frac{\langle \Psi(t)|H(t)|\Psi(t)\rangle}{\langle \Psi(t)|\Psi(t)\rangle}-E_{gs}&=\left[\sum_{q>0}vq|\alpha_{q}|^{2}+\sum_{q>0}2ig\mathcal{R}e(\alpha_{q})\right]-\frac{vK h^{2}(t)L}{2\pi},
\end{align}
where $E_{gs}$ is the ground state energy of the final Hermitian Hamiltonian. Note that the norm is a coherent state overlap multiplied by a phase. However, the phase contributions of the numerator and denominator cancel each other. For the linear quench driving $h(t)=h_{0}t/\tau$, using $\omega(q)=vq$, $\alpha_{q}(t)$ simplifies to
\begin{align}
\alpha_{q}(t)&=\frac{h_{0}}{\tau}\sqrt{\frac{K}{\pi L}}\frac{(1-\cos qL)}{\sqrt{q}}\frac{-e^{-itvq}+(1-itvq)}{vq^{2}}.\label{eq:alpha_q}
\end{align}
One can now express the general expression for the mean energy. Let us express the different terms separately
\begin{align}
\sum_{q>0}vq|\alpha_{q}(t)|^{2}&=\sum_{q>0}vq\left(\frac{h_{0}}{\tau}\right)^{2}\frac{K}{\pi L}\frac{(1-\cos qL)^{2}}{q}\frac{1}{(vq^{2})^{2}}\big[(1-\cos tvq)^{2}+(tvq-\sin tvq)^{2}\big]\\
&=\left(\frac{h_{0}}{\tau}\right)^{2}\frac{K}{\pi L}\frac{1}{v}\left(\frac{L}{\pi}\right)^{4}\sum_{n=1}^{\infty}\frac{(1-\cos  n\pi)^{2}\big[(1-\cos \frac{tv n\pi}{L})^{2}+(\frac{n\pi tv}{L}-\sin \frac{tvn\pi}{L})^{2}\big]}{n^{4}}\\
&=\left(\frac{h_{0}}{\tau}\right)^{2}\frac{K}{\pi L}\frac{1}{v}\left(\frac{L}{\pi}\right)^{4}\left[-\frac{4\pi tv}{L}\sum_{\sigma=\pm}\mathcal{I}m\text{Li}_{3}(\sigma e^{\frac{\sigma i\pi t v}{L}})-4\sum_{\sigma=\pm}\sigma \mathcal{R}e\text{Li}_{4}(\sigma e^{\frac{i\pi t v}{L}})+\frac{\pi^{4}}{12}+\frac{\pi^{4}t^{2}v^{2}}{2L^{2}}\right].
\end{align}
The other term reads
\begin{align}
\sum_{q>0}g\mathcal{R}e[\alpha_{q}(t)]&=\frac{K v}{\pi L}\frac{h_{0}t}{\tau}\frac{h_{0}}{\tau}\sum_{q>0}\frac{(1-\cos qL)^{2}}{q}\frac{1-\cos tvq}{vq^{2}}\\
&=\frac{K }{\pi L}\frac{h_{0}t}{\tau}\frac{h_{0}}{\tau}\left(\frac{L}{\pi}\right)^{3}\sum_{n=1}^{\infty}(1-\cos n\pi)^{2}\frac{1-\cos \frac{tvn\pi}{L}}{n^{3}}\\
&=\frac{K}{\pi L}\frac{h^{2}_{0}t}{\tau^{2}}\left(\frac{L}{\pi}\right)^{3}\left[\frac{7}{2}\zeta(3)-2\sum_{\sigma=\pm}\sigma\mathcal{R}e\text{Li}_{3}(\sigma e^{\frac{i\pi t v}{L}})\right].
\end{align}
Finally, the total energy is given by
\begin{align}
E({t})&=\left(\frac{h_{0}}{\tau}\right)^{2}\frac{K}{\pi^{2} v}\left(\frac{L}{\pi}\right)^{3}\Big\{\left[-4\tilde{t}\sum_{\sigma=\pm}\mathcal{I}m\text{Li}_{3}(\sigma e^{i\sigma\tilde{t}})-4\sum_{\sigma=\pm}\sigma \mathcal{R}e\text{Li}_{4}(\sigma e^{i\tilde{t}})+\frac{\pi^{4}}{12}+\frac{\pi^{2}}{2}\tilde{t}^{2}\right]\nonumber\\
&+2i\tilde{t}\left[\frac{7}{2}\zeta(3)-2\sum_{\sigma=\pm}\sigma\mathcal{R}e\text{Li}_{3}(\sigma e^{i\tilde{t}})\right]-\frac{\pi^{2}}{2}\tilde{t}^{2}\Big\}\label{transitory_energy},
\end{align}
with the dimensionless time $\tilde{t}=\pi t v/L$ for $t=\tau$, one recovers the expression in the main text ({\color{RubineRed}$12$}). It is plotted in Fig. \ref{fig:transit}.
\section{Field expectation values}
In order to compute the Imbalance, one needs the expectation value of the field $\phi(x)$. Using the decomposition of the field in the bosonic modes basis, one can evaluate the expression for a linear quench $h(t)=h_{0}t/\tau$, in using the expression for $\alpha_{q}(t)$ ({\color{RubineRed}$37$})
\begin{align}
\frac{\langle \Psi(t)|\phi(x)|\Psi(t)\rangle}{\langle \Psi(t)|\Psi(t)\rangle}&=-2\sqrt{\frac{\pi K}{L}}\sum_{q>0}\frac{\sin qx}{\sqrt{q}}\mathcal{I}m\ \alpha_{q}(t)\\
&=-\frac{2}{v}\frac{K}{L}\frac{h_{0}}{\tau}\sum_{q>0}\sin qx\frac{(1-\cos qL)}{q^{3}}[\sin tvq-tvq]\\
&=-\frac{2}{v}\frac{K}{L}\frac{h_{0}}{\tau}\left(\frac{L}{\pi}\right)^{3}\left[-\frac{\pi t v}{L}\sum_{\sigma=\pm}\sigma \mathcal{I}m \text{Li}_{2}(\sigma e^{\frac{i\pi x}{L}})-\frac{1}{2}\sum_{\beta,\sigma=\pm}\beta\sigma\mathcal{R}e\text{Li}_{3}(\sigma e^{\frac{i\pi(t v+\beta x)}{L}})\right].
\end{align}
From this on, one can directly obtain the expression of the asymptotic value given by
\begin{align}
\mathcal{I}(L/2,\infty)=8Kh_{0}\left(\frac{1}{\pi}\right)^{3}\left[\mathcal{I}m\text{Li}_{2}(i)-\mathcal{I}m\text{Li}_{2}(-i)\right],
\end{align}
leading to the expression of the main text ({\color{RubineRed}$16$}).
\section{Loschmidt echo}
The logarithm of the Loschmidt echo is given by ${\rm ln}\mathcal{L}(t)=-\sum_{q>0} |\alpha_{q}(t)|^{2}$ that can be expressed analytically for the linear quench $h(t)=h_{0}t/\tau$
\begin{align}
{\rm ln}\mathcal{L}(\tau)&=-\left(\frac{h_{0}}{\tau}\right)^{2}\frac{K}{\pi v^{2} L }\sum_{q>0}\frac{(1-\cos qL)^{2}}{q^{5}}\big[(1-\cos \tau vq)^{2}+(\sin \tau vq-\tau vq)^{2}\big]\\
&=-\left(\frac{h_{0}}{\tau}\right)^{2}\frac{K}{\pi v^{2} L }\left(\frac{L}{\pi}\right)^{5}\left[-\frac{4\pi \tau v}{L}\sum_{\sigma=\pm}\sigma \mathcal{I}m \text{Li}_{4}\sigma e^{\frac{i\pi \tau v}{L}}-4\sum_{\sigma=\pm}\sigma\mathcal{R}e\text{Li}_{5}\sigma e^{\frac{i\pi \tau v}{L}}+\frac{1}{4L^{2}}(14\pi^{2}\tau^{2}v^{2}\zeta(3)+31 L^{2}\zeta(5))\right].
\end{align}
In the large time limit, the amplitude of the Loschmidt echo decays in $\tau^{-1}$ with oscillations of period $2L/v$ towards an asymptotic value ${\rm ln\mathcal{L}(\infty)}=-\frac{7}{2}\frac{h^{2}_{0}KL^{2}}{\pi^{4}}\zeta(3)$, dividing by the asymptotic value leads to the expression in the main text ({\color{RubineRed}$18$}).

\section{Nonlinear quenches}

Nonlinear power-law quenches, with $h(t)=h_0 (t/\tau)^r$, lead to
\begin{eqnarray}
\alpha_q(t)&=&g_q(\tau)\left(\frac{t}{\tau}\right)^r\frac{e^{-i\omega(q)t}t\left[\Gamma(r+1)-\Gamma(r+1,-i\omega(q)t)\right]}{(-it\omega(q))^{r+1}}\\
&=&\frac{g_q(\tau)}{\omega(q)(\omega(q)\tau)^r}i^{r+1}e^{-i\omega(q)t}\left[\Gamma(r+1)-\Gamma(r+1,-i\omega(q)t)\right],%\label{exact_alpha}
\end{eqnarray}
in terms of the gamma and incomplete gamma functions, $\Gamma(z)$ and $\Gamma(a,z)$.
To simplify this expression it is convenient to resort to an asymptotic expansion in $1/t$ or to assume that $r$ is a positive integer.
In the latter case, one can use the identity
\begin{equation}
 \Gamma(r,z)=(r-1)!e^{-z}\sum_{k=0}^{r-1}\frac{z^k}{k!},   
\end{equation}
to find
\begin{eqnarray}
\alpha_q(t)&=&
\frac{g_q(\tau)}{\omega(q)(\omega(q)\tau)^r}i^{r+1}r!\sum_{k=r+1}^\infty\frac{(-i\omega(q)t)^k}{k!}\\
&=&\frac{g_q(\tau)}{\omega(q)(\omega(q)\tau)^r}i^{r+1}e^{-i\omega(q)t}r!\left[1-e^{i\omega(q)t}\sum_{k=0}^r\frac{(-i\omega(q)t)^k}{k!}\right]\\
&=&
\frac{g_{q}(\tau)}{\omega(q)(\omega(q)\tau)^{r}}i^{r+1}r!\left[e^{-i\omega(q)t}-1+\sum_{m=1}^{s}\frac{(\omega(q)t)^{2m}}{(2m)!}+i\left(\omega(q)t-\sum_{m=1}^{s'}\frac{(\omega(q)t)^{2m+1}}{(2m+1)!}\right)\right],
\end{eqnarray}
with $s'=r/2-1=s-1$ for even $r=2s$ and $s'=r/2=s$ for odd  $r=2s+1$.
For compactness, we write
\begin{eqnarray}
\alpha_q(t)&=&\frac{g_q(\tau)}{\omega(q)(\omega(q)\tau)^r}i^{r+1}r!\left[e^{-i\omega(q)t}+f_e(q,t)-if_o(q,t)\right],
\end{eqnarray}
such that
\begin{eqnarray}
|\alpha_q(t)|^2
%&=&K\left(\frac{h_0vr!}{\omega(q)(\omega(q)\tau)^r}\right)^{2}\frac{[1-\cos(qL)]^2}{\pi L q}\left[1+f_e(q,t)^2+f_o(q,t)^2+2f_e(q,t)\cos[\omega(q)t]+2f_o(q,t)\sin[\omega(q)t]\right],\\
&=&K\left(\frac{h_0vr!}{\omega(q)(\omega(q)\tau)^r}\right)^{2}\frac{[1-\cos(qL)]^2}{\pi L q}\left[
(f_e(q,t)+\cos[\omega(q)t])^2+(f_o(q,t)+\sin[\omega(q)t])^2\right].
\end{eqnarray}

Consider
\begin{eqnarray}
\sum_{q\geq 0}vq|\alpha_q(t)|^2
&=&vK\left(\frac{h_0r!}{(v\tau)^r}\right)^{2}\sum_{n=1}^\infty \left(\frac{L}{n\pi}\right)^{1+2r}\frac{[1-\cos(n\pi)]^2}{n\pi^2 }\left[
\left(f_e\left(\frac{n\pi}{L},t\right)+\cos\frac{tvn\pi}{L}\right)^2+\left(f_o\left(\frac{n\pi}{L},t\right)+\sin\frac{tvn\pi}{L}\right)^2\right].\nonumber\\
& =&vK\left(\frac{h_0r!}{(v\tau)^r}\right)^{2}\sum_{n=1}^\infty \left(\frac{L}{n\pi}\right)^{1+2r}\frac{[1-\cos(n\pi)]^2}{n\pi^2 }\left[
\left(f_e(n,\tilde{t})+\cos n \tilde{t} \right)^2+\left(f_o(n,\tilde{t})+\sin\tilde{t}\right)^2\right],
\end{eqnarray}
with $\tilde{t}=tv\pi/L$, $f_e(n,\tilde{t})=-1+\sum_{m=1}^s\frac{(n\tilde{t})^{2m}}{(2m)!}$, $f_o(n,\tilde{t})=-n\tilde{t}+\sum_{m=1}^{s'}\frac{(n\tilde{t})^{2m+1}}{(2m+1)!}$.

In general, this expression cannot be summed in closed form but is amenable to a numerical evaluation.

We further note that the short-time expansion
\begin{eqnarray}
\alpha_q(t)&=&
-\frac{g_q(\tau)}{\omega(q)(\omega(q)\tau)^r}i^{r+1}r!\left[1-e^{-i\omega(q)t}-i\omega(q)t+\mathcal{O}(t^2)\right]
\end{eqnarray}
takes a functional form reminiscent of that in the linear quench.
Thus, expressions obtained in the linear case generalize to the non-linear quenches at short times by replacing
\begin{eqnarray}
\frac{1}{\omega(q)^2\tau}\rightarrow 
\frac{-i^{r+1}r!}{\omega(q)(\omega(q)\tau)^r}.
\end{eqnarray}

It then follows that
\begin{eqnarray}
|\alpha_q(t)|^2
%&=&K\left(\frac{h_0vr!}{\omega(q)(\omega(q)\tau)^r}\right)^{2}\frac{[1-\cos(qL)]^2}{\pi L q}\left[1+f_e(q,t)^2+f_o(q,t)^2+2f_e(q,t)\cos[\omega(q)t]+2f_o(q,t)\sin[\omega(q)t]\right],\\
&\approx&K\left(\frac{h_0vr!}{\omega(q)(\omega(q)\tau)^r}\right)^{2}\frac{[1-\cos(qL)]^2}{\pi L q}\big[(1-\cos tvq)^{2}+(tvq-\sin tvq)^{2}\big].
\end{eqnarray}
This expression makes it possible to compute approximately the energy, imbalance, and Loschmidt echo in close form. 

For instance,
\begin{eqnarray}
{\rm ln}\mathcal{L}(t)
&=&
-\left(\frac{h_0r!}{v^r\tau^r}\right)^{2}\frac{K}{\pi L}\sum_{q>0}\frac{[1-\cos(qL)]^2}{q^{3+2r}}\frac{\left[\Gamma(r+1)-\Gamma(r+1,-iv qt)\right]\left[\Gamma(r+1)-\Gamma(r+1,+iv qt)\right]}{\Gamma(r+1)^2}\\
&\approx&
-\left(\frac{h_0r!}{v^r\tau^r}\right)^{2}\frac{K}{\pi L}\sum_{q>0}\frac{[1-\cos(qL)]^2}{q^{3+2r}}\big[(1-\cos tvq)^{2}+(tvq-\sin tvq)^{2}\big].
%\\
%&=&\left[-\frac{4\pi \tau v}{L}\sum_{\sigma=\pm}\sigma \mathcal{I}m \text{Li}_{4}\sigma e^{\frac{i\pi \tau v}{L}}-4\sum_{\sigma=\pm}\sigma\mathcal{R}e\text{Li}_{5}\sigma e^{\frac{i\pi \tau v}{L}}+\frac{1}{4L^{2}}(14\pi^{2}\tau^{2}v^{2}\zeta(3)+(8-2^{-r3}) L^{2}\zeta(3+2r))\right].
\end{eqnarray}

For instance, for $r=2$, one obtains the expression
\begin{align}
\label{LEr2approx}
{\rm ln}\mathcal{L}(t) \approx
-\left(\frac{h_02}{v^2\tau^2}\right)^{2}\frac{K}{\pi L}
\left(\frac{L}{\pi}\right)^{7}
\bigg(\sum_{\sigma=\pm}\sigma\left[-4\mathcal{R}e\text{Li}_{7}(\sigma e^{i\tilde{t}}) -4\tilde{t} \mathcal{I}m\text{Li}_{6}(\sigma e^{i\tilde{t}}) \right]
   +\tilde{t}^2\frac{31}{8}\zeta(5)+\frac{127}{16}\zeta(7)\bigg).
\end{align}
For arbitrary $r$, one finds
\begin{eqnarray}
{\rm ln}\mathcal{L}(t) & \approx&
-\left(\frac{h_0r!}{v^r\tau^r}\right)^{2}\frac{K}{\pi L}
\left(\frac{L}{\pi}\right)^{3+2r}
\\
& &\bigg(\sum_{\sigma=\pm}\sigma\left[-4\mathcal{R}e\text{Li}_{3+2r}(\sigma e^{i\tilde{t}}) -4\tilde{t} \mathcal{I}m\text{Li}_{2+2r}(\sigma e^{i\tilde{t}}) \right]
    +(4-2^{1-2r})\tilde{t}^2\zeta(1+2r)+(8-2^{-2r})\zeta(3+2r)\bigg).\nonumber
\end{eqnarray}

Beyond $r=1$, the exact computation for arbitrary $t$ can be found for low specific values of $r$. For instance, for $r=2$, the exact Loschmidt echo is given by
\begin{eqnarray}
{\rm ln}\mathcal{L}(t) & =&
-\left(\frac{h_02}{v^2\tau^2}\right)^{2}\frac{K}{\pi L}
\left(\frac{L}{\pi}\right)^{7}
\\
& &\times\bigg(\sum_{\sigma=\pm}\sigma\left[-4\mathcal{R}e\text{Li}_{7}(\sigma e^{i\tilde{t}}) -4\tilde{t} \mathcal{I}m\text{Li}_{6}(\sigma e^{i\tilde{t}}) +2\tilde{t}^2\mathcal{R}e\text{Li}_{5}(\sigma e^{i\tilde{t}})\right]
   +\tilde{t}^4\frac{7}{8}\zeta(3)+\frac{127}{16}\zeta(7)\bigg),\nonumber
\end{eqnarray}
to be contrasted with the approximate form in Eq. ({\color{RubineRed}$65$}).

\section{Energy for the nonlinear quenches}
Following the procedure described in the main text, one can compute the bosonized expression for the residual energy after the quench at time $\tau$. We study the nonlinear power-law quench, with $h(t)=h_{0}(t/\tau)^{r}$. Let us consider the case $r=2$, for which we obtain
\begin{align}
E(\tau)-E_{gs}&=\frac{1}{30 \pi^{7}v^{3}\tau^{4} }\Big(h^{2}_{0}K L^{5}\pi^{6}+15 h^{2}_{0}K L \pi^{6}v^{4}\tau^{4}+240 h^{2}_{0}K L^{3}\pi^{2}v^{2}\tau^{2}\sum_{\sigma=\pm}\sigma \mathcal{R}e\text{Li}_{4}(\sigma e^{i\tilde{\tau}})\nonumber\\
&-480 h^{2}_{0}L^{4}\pi v \tau \sum_{\sigma=\pm}\sigma \mathcal{I}m\text{Li}_{5}(\sigma e^{i\tilde{\tau}})-480 h^{2}_{0}L^{5}K\sum_{\sigma=\pm}\sigma \mathcal{R}e\text{Li}_{6}(\sigma e^{i\tilde{\tau}})\Big)\nonumber\\
&-\frac{8i}{\pi^{5} v\tau^{2} }h^{2}_{0}K L^{3}\sum_{\sigma=\pm}\sigma \mathcal{I}m\text{Li}_{4}(\sigma e^{i\tilde{\tau}})+\frac{14 i}{\pi^{5}v\tau^{2}}h^{2}_{0}KL^{2}\pi v \tau \zeta(3).
\end{align}
Also, consider the case $r=3$
\begin{align}
E(\tau)-E_{gs}&=\frac{1}{560 \pi^{9}v^{5}\tau^{6}}h^{2}KL\Big(17 L^{6}\pi^{8}-70 L^{2}\pi^{8}v^{4}\tau^{4}+280\pi^{8}v^{6}\tau^{6}-13440 i L^{3}\pi^{3}v^{3}\tau^{3}\text{Li}_{5}(-e^{-i\tilde{\tau}})\\
&+13440 i L^{3}\pi^{3}v^{3}\tau^{3}\text{Li}_{5}(e^{-i\tilde{\tau}})+40320 L^{4}\pi^{2}v^{2}\tau^{2}\sum_{\sigma=\pm}\sigma \mathcal{R}e\text{Li}_{6}
(\sigma e^{i\tilde{\tau}})-80640 L^{5}\pi v\tau \sum_{\sigma=\pm}\mathcal{I}m\text{Li}_{7}(\sigma e^{\sigma i\tilde{\tau}})\\
&-80640 L^{6}\sum_{\sigma=\pm}\sigma\mathcal{R}e\text{Li}_{8}(\sigma e^{i\tilde{\tau}})+11760  i L \pi^{5}v^{5}\tau^{5}\zeta(3)-26040 i L^{3}\pi^{3}v^{3}\tau^{3}\zeta(5)\Big).
\end{align}
\begin{figure}
    \centering
    \includegraphics[width=0.9\linewidth]{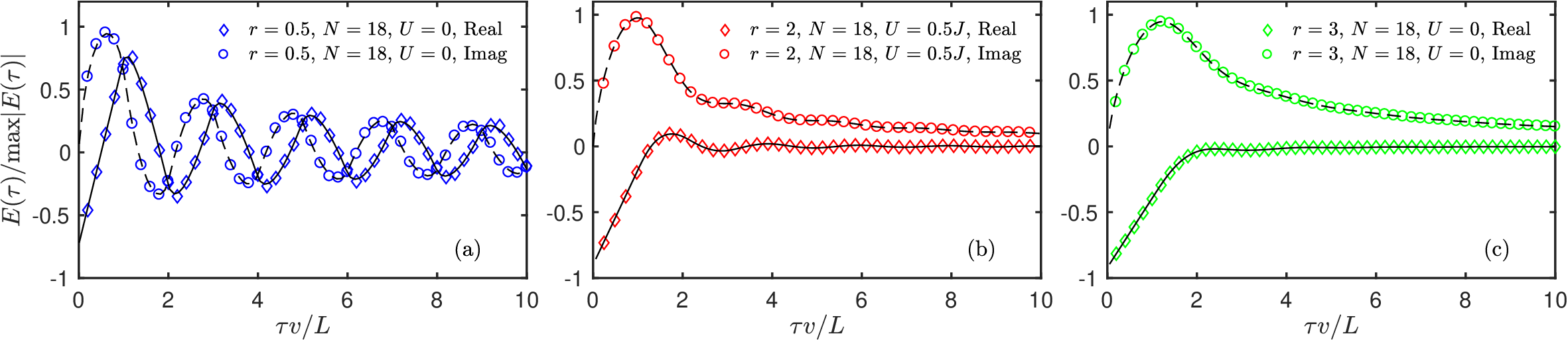}
    \caption{Evolution of the energy for power law quenches $h(t)=h_{0}(t/\tau)^{r}$, figure (a) corresponds to $r=0.5$, figure (b) to $r=2$ and figure (c) to $r=3$.  The numerical simulation is plotted in symbols and the bosonization in plain line for the real part and dashed line for the imaginary part. The case $r=0.5$ depicts strong oscillations with $\tau^{-1}$ amplitude decay. For $r=2$, one still observes small amplitude decaying oscillations, and for $r=3$ the oscillations vanish. }
    \label{fig5}
\end{figure}
Fig. \ref{fig5} displays the behavior of the residual energy after the quench for different nonlinear quenches $r=0.5$, $r=2$ and $r=3$. The value of the energy for $r=2$ and $r=3$ can be computed analytically in using bosonization, which is compared to numerical simulation. For $r=0.5$, an analytical expression is not found so that we compute the infinite sum numerically using Eq. ({\color{RubineRed}$11$}), and compare the result to numerical simulation. For quench $r=0.5$, one observes an oscillatory behavior for the real and imaginary part, with a decaying amplitude in $\tau^{-1}$. For the case $r=2$ the oscillations are damped and for $r=3$, one does not observe oscillations anymore.  

\end{document}